\documentclass[letterpaper,11pt,leqno]{article}
\usepackage{paper,math}
\hypersetup{pdftitle={Beveridgean Phillips Curve}}
\newcommand{\bib}{paper.bib}
\newcommand{\wpdf}{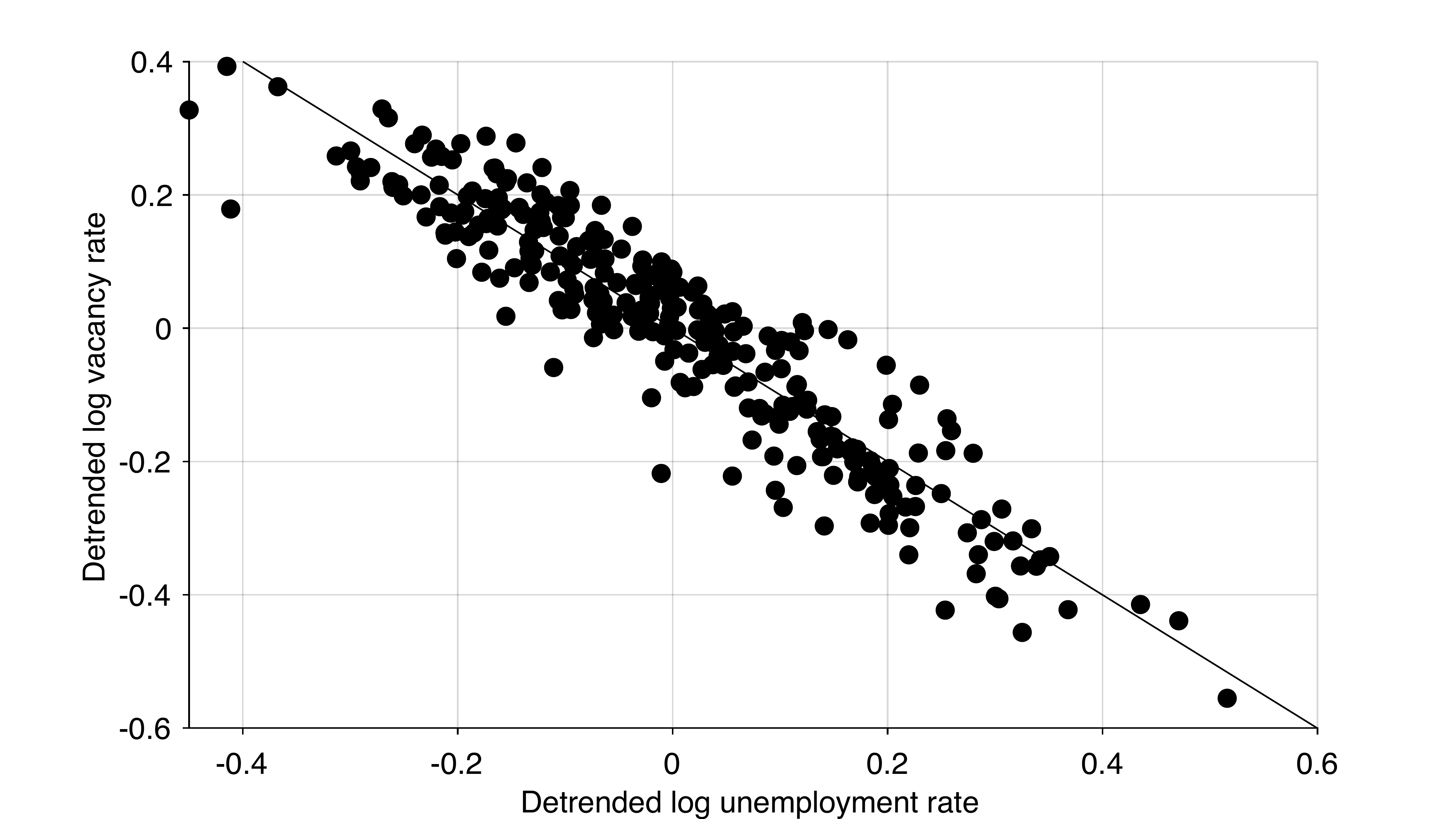}
\newcommand{\npdf}{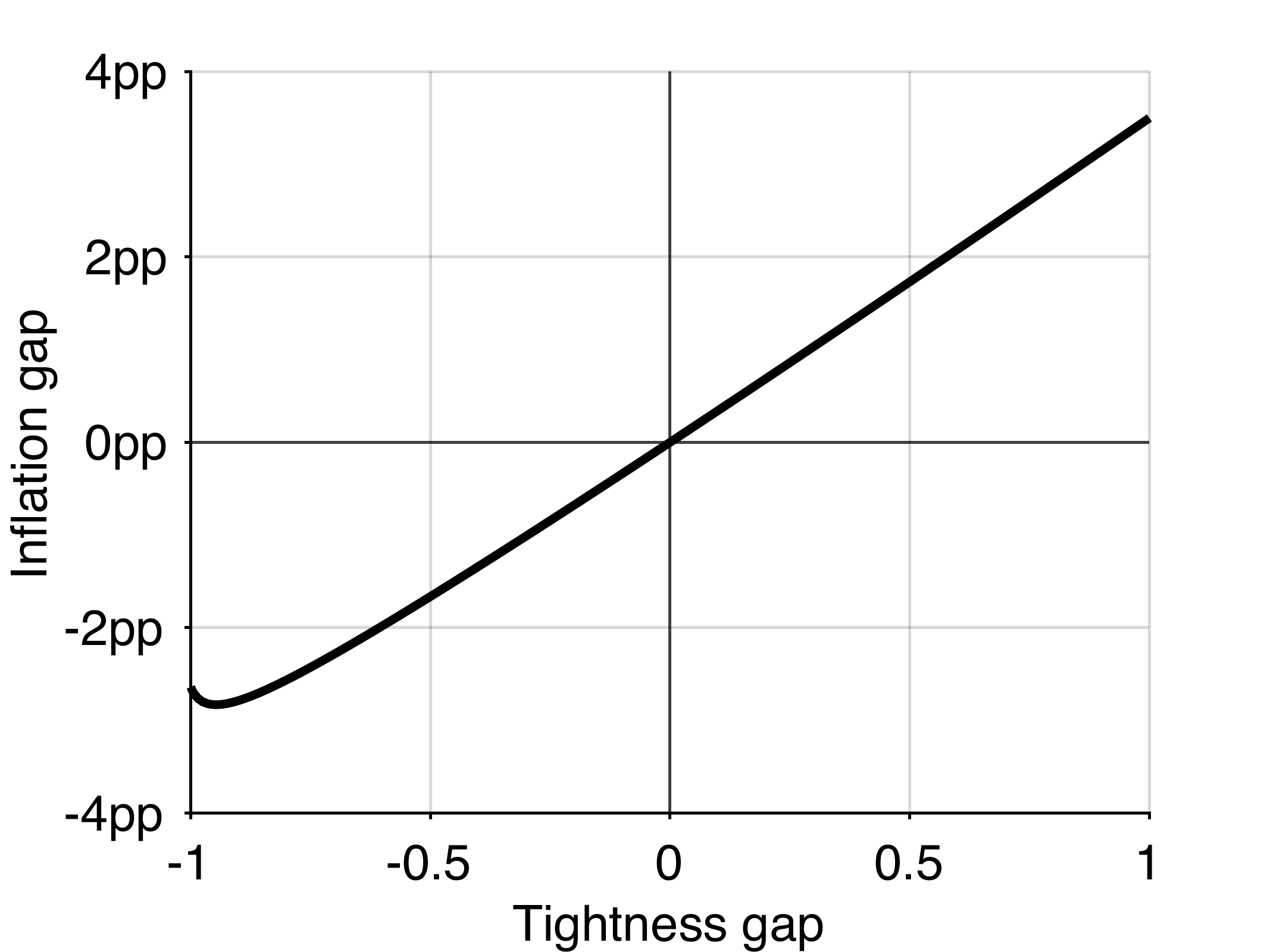}
\begin{document}

\title{Beveridgean Phillips Curve}
\author{Pascal Michaillat, Emmanuel Saez
\thanks{Michaillat: University of California--Santa Cruz. Saez: University of California--Berkeley. We thank Pierpaolo Benigno, Allegra Saggese, and Jon Steinsson for helpful comments and discussions.}}
\date{September 2026}
\paperurl{https://pascalmichaillat.org/15/} 
\begin{titlepage}\maketitle

This paper develops a Beveridgean model of the Phillips curve. While the New Keynesian Phillips curve is based on monopolistic pricing under price-adjustment costs, the Beveridgean Phillips curve is based on directed-search pricing under price-adjustment costs. Under directed search, prices respond to slack instead of marginal costs. The resulting Phillips curve has three properties that match recent US evidence. First, it delivers the divine coincidence: inflation is on target whenever unemployment is at its efficient level---the full-employment rate of unemployment (FERU), the geometric mean of the unemployment and vacancy rates. Second, it is decreasing and convex in the unemployment-inflation plane: inflation responds more strongly to unemployment when the labor market is inefficiently tight than when it is inefficiently slack. The convexity is inherited from the Beveridge curve, which makes unemployment a decreasing and convex function of tightness. Third, shifts of the Beveridge curve move the FERU and therefore shift the unemployment-inflation Phillips curve. The model also implies that the Fed's dual mandate is internally consistent, that a soft landing from an inefficiently tight labor market is possible, and that the zero lower bound is not topsy-turvy.

\end{titlepage}\section{Introduction}\label{s:introduction}

The post-pandemic spike in inflation in the United States, which coincided with wide labor-market fluctuations, revealed new facts about the Phillips curve. First, the Phillips curve ensures the divine coincidence \citep{BG07}: inflation is on target whenever the labor market is efficient; inflation rises above target whenever the labor market is inefficiently tight; and inflation falls below target whenever the labor market is inefficiently slack \citep{BE23,G24}. Second, inflation is driven by slack rather than marginal costs: labor market tightness---the ratio of job vacancies to unemployed workers---predicts inflation better than the unemployment rate or measures of marginal costs \citep{BOS21,BS22,BS24}. Third, the Phillips curve is not linear: it is much steeper when the labor market is inefficiently tight than when the labor market is inefficiently slack \citep{BD17,STW25,BE23}. These empirical findings are summarized by figure~\ref{f:pandemic}.

To explain these facts, we propose a Beveridgean model of the Phillips curve. The economy is a matching economy. Households purchase and consume services produced by other households, and all trades are mediated by a matching function---which generates unemployment, help-wanted ads, and a Beveridge curve \citep{PP01}. To generate price dynamics, we introduce price competition through directed search \citep{M97}. Each household posts help-wanted ads that offer a specific price for services, and jobseekers choose which ads to pursue, trading off the price offered against the time required to find a job. To ensure that slack fluctuates, we introduce price rigidity through quadratic price-adjustment costs \citep{R82}. In the model, directed search produces a Phillips curve linking labor-market slack to inflation, and the price-adjustment costs ensure that the Phillips curve is nonvertical.

The Phillips curve guarantees the divine coincidence: inflation is on target if and only if the labor market operates efficiently. The coincidence follows from two ingredients. The first is the efficiency of directed search \citep{M97}: absent price-adjustment costs, price-setters would drive labor market tightness to its efficient level, because at that level the resources lost to unemployment and recruiting are minimized. With price-adjustment costs, price-setters deviate from the efficient tightness only insofar as inflation deviates from the inflation that people find normal. The second ingredient is that the central bank anchors normal inflation at its inflation target---an assumption that is realistic for the modern United States, where long-run inflation expectations have been fixed slightly above 2\% for three decades \citep[figure~2]{HHN22}. Combining the two ingredients, inflation is on target when tightness is efficient and unemployment is at the full-employment rate of unemployment (FERU).

In the unemployment-inflation plane, the Phillips curve is convex: steep when unemployment is inefficiently low and flat when unemployment is inefficiently high. The reason is simple. Inflation is roughly linear in labor market tightness, but unemployment is a convex function of tightness through the Beveridge curve, so inflation is convex in unemployment. We prove this analytically when the matching function is Cobb-Douglas and symmetric, and we verify it numerically in the calibrated model. The convexity also explains why empirical studies detect a kink in the Phillips curve at an unemployment rate of 4.2\% \citep{BD17,STW25}---the average FERU in the postwar United States---since a convex curve looks kinked in finite samples. Existing models instead generate convexity or a kink from constraints on wage movements: nominal wages that are downwardly rigid \citep{ADP96,BR11a,CFG03}, or downwardly rigid only when the labor market is slack \citep{BE23}.

The model also explains why the unemployment-inflation Phillips curve is unstable while its tightness formulation is not. Whenever the Beveridge curve shifts, the FERU moves, so the point of divine coincidence travels in the unemployment-inflation plane and the Phillips curve shifts. Beveridge shocks therefore microfound the cost-push shocks of the New Keynesian literature. The US Beveridge curve shifted dramatically outward when the pandemic hit, pushing the FERU from 4.0\% to 6.7\% in one quarter; the FERU then slowly came back down. The efficient tightness, by contrast, is pinned down by two stable sufficient statistics---the Beveridge elasticity and the recruiting cost---so it barely moves when the Beveridge curve shifts. This is why the tightness-inflation Phillips curve performed so much better than the unemployment-inflation Phillips curve during the pandemic \citep{BOS21,BS22,BS24}.

The divine coincidence has several implications for monetary policy. The Fed's full-employment and price-stability mandates coincide: by maintaining the economy at full employment, the Fed automatically keeps inflation on target. The interest rate that achieves both mandates is the full-employment rate of interest (FERI). From an inefficiently tight labor market, it is also possible for the Fed to achieve a soft landing \citep{B23}. Cooling the labor market from its 2022 peak (tightness of 2) to efficiency (tightness of 1) raises unemployment only from 3.5\% to 5.0\% along a frozen Beveridge curve---and by less if matching efficacy is still improving. The convexity of the Phillips curve further suggests a state-contingent choice of policy targets: the Fed should target the FERU when the economy is inefficiently slack and the inflation target when the economy is inefficiently tight. Finally, the zero lower bound is not the topsy-turvy world of the New Keynesian model \citep{EK12}: at the zero lower bound the model admits a unique solution that behaves like a normal slump, even if the bound is permanent.

\section{Slack and full employment in the United States}\label{s:slack}

Before developing the model, we present how we measure slack, full employment, and the unemployment gap in the United States. The measurement framework is borrowed from \citet{MS24}.

\subsection{Unemployment, vacancies, and tightness}

We measure labor market slack from three series covering 1929--2024: the unemployment rate, the vacancy rate, and their ratio, labor market tightness.

The unemployment rate $u$ is the number of jobseekers as a share of the labor force. Between 1948 and 2024, it is computed from the Current Population Survey \citep{UNEMPLOY,CLF16OV}. Between 1929 and 1947, it comes from the historical series constructed by \citet{PZ21}. The unemployment rate is sharply countercyclical. Over 1929--2024, it averages 6.4\%; over the more stable 1948--2019 period, it averages 5.7\%.

The vacancy rate $v$ is the number of job openings as a share of the labor force. Between 2001 and 2024, it is computed from the Job Openings and Labor Turnover Survey \citep{JTSJOL,CLF16OV}. Between 1951 and 2000, it comes from the vacancy proxy constructed by \citet{B10} from the Conference Board's help-wanted advertising index. Between 1929 and 1950, it comes from the vacancy proxy constructed by \citet{PZ21} from MetLife's help-wanted index. The vacancy rate is sharply procyclical. Over 1929--2024, it averages 3.2\%.

Labor market tightness is the ratio of the two rates: $\t = v/u$. It is the number of vacant jobs per jobseeker. Tightness is a natural measure of slack in any matching model: it determines how quickly jobseekers find jobs and how quickly employers fill vacancies.

Because unemployment is countercyclical and vacancies are procyclical, the two rates are negatively related: they trace out a \citet{B44} curve. In the United States, the Beveridge curve is close to a rectangular hyperbola, $uv = A$. When the log vacancy rate is regressed on the log unemployment rate, after removing a slow-moving trend from both series to absorb the decennial shifts of the curve, the estimated elasticity over 1948--2019 is $-1.00$, with an $R^2$ of 0.87 (figure~\ref{f:hyperbola}).

\begin{figure}[t]
\widegraphic[1]{\wpdf}
\caption{The US Beveridge curve is a rectangular hyperbola}
\note{The unemployment and vacancy rates come from the sources listed in section~\ref{s:slack}. The figure plots the logarithm of the vacancy rate against the logarithm of the unemployment rate, both detrended using a HP filter with smoothing parameter 10,000. The data are quarterly, 1948--2019. The regression line has a slope of $-1.00$, indicating that the US Beveridge curve is isoelastic with elasticity $\b = 1$: a rectangular hyperbola.}
\label{f:hyperbola}\end{figure}

\subsection{Efficient tightness and the FERU}\label{s:feru}

Slack is inefficient when it is excessive, but some slack is unavoidable, and even desirable. Reducing unemployment requires posting more vacancies, and vacancies absorb labor. Indeed, recruiting takes work: designing and advertising job vacancies, screening and interviewing candidates, negotiating contracts. 

The number of producers in the economy---workers who are neither unemployed nor recruiting---is therefore $(1 - u - \k v) h$, where $h$ is the labor force. Efficiency requires choosing the point on the Beveridge curve $v(u)$ that maximizes the number of producers, which means minimizing $u + \k v(u)$. The solution takes a simple sufficient-statistic form \citep{MS21b}:
\begin{equation}
\t^* = \frac{1}{\k\b},
\label{e:thetastar}\end{equation}
where $\b = -\oex{v}{u}$ is the elasticity of the Beveridge curve.

In the United States, servicing one vacancy requires about one full-time worker: estimates from the National Employer Survey imply 0.92 workers per vacancy \citep[p.~11]{MS21b}, and estimates from a survey by Bersin \& Associates imply 1.16 workers per vacancy \citep[p.~332]{MS24}. Hence the number of vacancies effectively measures the number of recruiters, and the recruiting cost is $\k = 1$ worker per vacancy.

With $\k = 1$ and $\b = 1$ in US data, the efficient tightness is
\begin{equation*}
\t^* = 1.
\end{equation*}
The US labor market is inefficiently slack whenever tightness is below 1, inefficiently tight whenever tightness is above 1, and efficient---at full employment---when there are as many vacant jobs as jobseekers.

The unemployment rate that prevails at the efficient tightness is the full-employment rate of unemployment (FERU), denoted $u^*$. Because the Beveridge curve is a rectangular hyperbola, $uv = A$, and full employment requires $v^* = u^*$, the FERU satisfies $(u^*)^2 = A = uv$, so the FERU is the geometric mean of the unemployment and vacancy rates \citep{MS24}:
\begin{equation}
u^* = \sqrt{uv}.
\label{e:feru}\end{equation}
The FERU is strictly positive: full employment is not zero unemployment. Zero unemployment is infeasible---the Beveridge curve bounds unemployment away from zero---and undesirable---pushing unemployment toward zero would divert a vast amount of labor toward recruiting.

\subsection{A century of slack}

\begin{figure}[t]
\widegraphic[2]{\wpdf}
\caption{Labor market tightness in the United States, 1929--2024}
\note{Labor market tightness is $\t = v/u$, where the unemployment and vacancy rates come from the sources listed in section~\ref{s:slack}. Shaded areas indicate recessions dated by the \citet{NBER23}. The labor market is at full employment when tightness equals 1, inefficiently slack when tightness is below 1, and inefficiently tight when tightness is above 1.}
\label{f:tightness}\end{figure}

Figure~\ref{f:tightness} displays US labor market tightness over 1929--2024, along with the full-employment level of 1. Over the period, tightness averages 0.73 and generally remains below 1: the US labor market is typically inefficiently slack, especially in recessions. Tightness plunged to 0.03 during the Great Depression and climbed to 6.8 at the end of World War 2. The postwar period is quieter: tightness peaked at 1.6 in 1953:Q1, during the Korean War, and bottomed at 0.16 in 2009:Q3, during the Great Recession.

The labor market is not always inefficiently slack, however. Before 2018, it had been inefficiently tight only during major wars: World War 2, the Korean War, and the Vietnam War. But since 2018, the labor market has been inefficiently tight twice in peacetime: just before the coronavirus pandemic (2018:Q3--2020:Q1) and in its aftermath (2021:Q3--2024:Q4). In 2022:Q2, tightness reached 2.0, a value last seen in 1945. The years around the pandemic are therefore a rarity: the only episode of peacetime overheating on record.\footnote{People at the time were aware of the overheating. Processing about 20,000 news articles per month from major US newspapers, \citet[figure~2]{CIY26} construct an index of shortages in the United States from 1900 to 2024. The episodes of labor shortage that they detect overlap with the episodes of inefficient tightness in figure~\ref{f:tightness}: World War 2, the Korean War, the Vietnam War, and the pandemic era.}

\begin{figure}[p]
\subcaptionbox{Full-employment rate of unemployment (FERU)\label{f:feru}}{\widegraphic[3]{\wpdf}}\fspace
\subcaptionbox{Unemployment gap\label{f:gap}}{\widegraphic[4]{\wpdf}}
\caption{FERU and unemployment gap in the United States, 1929--2024}
\note{The unemployment and vacancy rates come from the sources listed in section~\ref{s:slack}. The FERU is $u^* = \sqrt{uv}$, so on a logarithmic scale it is the midpoint between the unemployment and vacancy rates. The unemployment gap is $u - u^*$. Shaded areas indicate recessions dated by the \citet{NBER23}.}
\label{f:fullemployment}\end{figure}

Figure~\ref{f:feru} displays the FERU computed from \eqref{e:feru}. Over 1929--2024, the FERU averages 4.1\%, and it is quite stable: it remains between 2.3\% and 6.7\% over almost a century. During the postwar period, 1948--2019, the FERU averages 4.2\% and remains between 3.1\% and 5.5\%. The Beveridge curve shifts in and out over the decades, but the shifts are usually too small to move the FERU much. The exceptions are the three major catastrophes of the century, which shifted the Beveridge curve sharply outward and raised the FERU abruptly: the Great Depression (FERU up 2.3pp between 1929:Q4 and 1930:Q4), the demobilization after World War 2 (up 1.7pp between 1944:Q4 and 1945:Q4), and above all the coronavirus pandemic. When the US economy shut down in the spring of 2020, the FERU jumped from 4.0\% in 2020:Q1 to 6.7\% in 2020:Q2---an increase of 2.7pp in one quarter. The FERU then slowly came back down, falling below 5\% in 2022 and below 4.5\% in 2023.

Figure~\ref{f:gap} displays the unemployment gap, $u - u^*$: the distance of the labor market from full employment. Over 1929--2024, the gap averages $+2.3$pp, confirming that the US labor market is generally inefficiently slack. The gap reached $+20.9$pp in 1932:Q3, at the depth of the Great Depression, and $-1.6$pp in 1945:Q1, at the peak of the war effort. In the postwar period, the gap peaked at $+5.9$pp in 2009:Q4, during the Great Recession. During the pandemic, the labor market experienced sharp departures from full employment in both directions. The gap first spiked to $+6.4$pp in 2020:Q2, its highest level since 1945. The labor market then recovered quickly, reached full employment in the middle of 2021, and overshot: the gap fell to $-1.5$pp in 2022:Q2, its lowest level since 1945, before shrinking back to $-0.2$pp in 2024:Q4.

\section{Three facts about the Phillips curve}\label{s:facts}

With the measurement framework in place, we turn to three facts about the Phillips curve in the United States.

\begin{figure}[p]
\subcaptionbox{Time series\label{f:time}}{\widegraphic[6]{\wpdf}}\fspace
\subcaptionbox{Scatter plot\label{f:scatter}}{\widegraphic[7]{\wpdf}}
\caption{Phillips curve in the United States, 2020--2024}
\note{The unemployment gap comes from figure~\ref{f:gap}. The inflation rate is the quarterly average of the percentage change from a year ago of the monthly, seasonally adjusted personal consumption expenditures price index, which is produced by the \citet{PCEPI}. The inflation gap is the inflation rate minus the inflation target of 2\%. These are the inflation measure and inflation target used by the Federal Open Market Committee \citep{FOMC26}.}
\label{f:pandemic}\end{figure} 

\subsection{First fact: divine coincidence}\label{s:fact1}

The first fact is that inflation is on target whenever the labor market is at full employment---the divine coincidence.

Figure~\ref{f:time} illustrates the parallel trajectories followed by the inflation gap and the unemployment gap during the pandemic and after (2020--2024). Inflation is measured as the annual change in the price index for personal consumption expenditures \citep{PCEPI}; the inflation gap is inflation minus the 2\% target \citep{FOMC26}. The unemployment gap comes from figure~\ref{f:gap}.

The two gaps evolve in tandem. Both are close to 0 at the beginning of 2020: the unemployment gap is $-0.2$pp and the inflation gap is $-0.4$pp---the economy is close to the point of divine coincidence. The following quarter, the unemployment gap spikes to $+6.4$pp while the inflation gap bottoms at $-1.4$pp. Both gaps then recover, returning to 0 in the first half of 2021. The inflation gap keeps rising and the unemployment gap keeps falling until 2022:Q2, when the inflation gap peaks at $+4.9$pp and the unemployment gap bottoms at $-1.5$pp. After that, both gaps normalize: by 2024:Q3, the inflation gap is down to $+0.4$pp and the unemployment gap up to $-0.1$pp. The economy is back at full employment and price stability.

This parallel pattern suggests that the divine coincidence prevails in the modern US economy: the inflation gap is 0 whenever the unemployment gap is 0. The coincidence occurred at the beginning of 2020, in the first half of 2021, and at the end of 2024. The two gaps are also negatively related: inflation is above target whenever unemployment is below the FERU, and below target whenever unemployment is above the FERU.

This basic evidence is supported by other work. Using aggregate US data over 2008--2022, \citet[figure~4]{BE23} find that when labor market tightness is 1, inflation is on target at 2\%. Using metropolitan-level US data over 2001--2022, \citet[figure~1]{G24} finds the same pattern, in noisier form: when local tightness is 1, local inflation is around 2\%. Before the pandemic, \citet{C17a} argued that the divine coincidence held in the United States. The idea has a long policy pedigree: in the 1960s, the Council of Economic Advisers believed that at full employment the economy was not subject to inflationary pressures \citep[p.~18]{B22}; and over the years the Fed indicated through its FOMC statements that it saw its two mandates as equivalent \citep[pp.~119--121, p.~130]{T12}.

\subsection{Second fact: slack drives inflation}\label{s:fact2}

The second fact is that inflation is driven by labor market slack, and that slack is best measured by labor market tightness.

\citet{BOS21} and \citet{BS22,BS24} show that a Phillips curve formulated with labor market tightness performs better than the traditional Phillips curve formulated with the unemployment rate: incorporating tightness improves inflation forecasts relative to unemployment alone, and relative to standard marginal-cost proxies. The superiority of the tightness formulation is especially visible in the pandemic period, when the Beveridge curve was unstable. Relatedly, \citet{BLM22} argue that the post-pandemic spike in inflation was so severe because labor market tightness was so elevated.

The tight connection between slack and inflation is visible in figure~\ref{f:pandemic}: between 2020 and 2024, the inflation gap tracks the unemployment gap---and therefore labor market tightness---closely at every phase of the cycle: the collapse in 2020, the overheating of 2021--2022, and the normalization of 2023--2024.

\subsection{Third fact: the Phillips curve is convex}\label{s:fact3}

The third fact is that the Phillips curve is much steeper when the labor market is inefficiently tight than when it is inefficiently slack.

The convexity appears clearly in the scatter plot of figure~\ref{f:scatter}. In early 2020, the economy was inefficiently slack, and most of the adjustment occurred through unemployment: the unemployment gap spiked to $+6.4$pp while the inflation gap fell only to $-1.4$pp. Once the economy became inefficiently tight after 2021, most of the adjustment occurred through inflation: the inflation gap spiked to $+4.9$pp while the unemployment gap fell only to $-1.5$pp. The Phillips curve is thus flat to the right of the point of divine coincidence and steep to the left of it.

More systematic evidence points to the same pattern. \citet[table~4]{BD17} and \citet[table~5]{STW25} find that inflation is much more responsive to unemployment along the US Phillips curve when the unemployment rate is below 4.2\% than when it is above 4.2\%. Similar nonlinearity is found in metropolitan US data \citep{BD17,STW25,G24} and in international data \citep{STW25,BE24a}. The kink is detected at an unemployment rate of 4.2\%---exactly the average FERU in the postwar United States (figure~\ref{f:feru}). These studies describe the pattern as a kink, whereas the Phillips curve developed in this paper is smooth but strictly convex; the two descriptions are hard to distinguish in finite samples, because a convex Phillips curve looks kinked.\footnote{\citet[figure~4]{BE23} also find that inflation is more responsive to tightness when tightness is above 1 than when it is below 1. The model developed here does not produce this property: inflation is essentially linear in tightness (figure~\ref{f:pctightness}).}

\section{Beveridgean model of the Phillips curve}\label{s:model}

We now develop the model, which adds a Phillips curve to the business cycle model developed in \citet{MS22}. In a New Keynesian model, price dynamics are driven by marginal costs.\footnote{See for instance \citet[equation~(3)]{GG99} and \citet[equation~(11)]{GGL25}.} Here, by contrast, price dynamics are driven by the amount of slack, in the tradition of \citet{P58} and of the early matching models of \citet{H70} and \citet{M70}. When the economy is inefficiently slack, buyers of services are pushed to lower the prices they pay. When the economy is inefficiently tight, they are pushed to raise them.\footnote{In New Keynesian models with a matching labor market, marginal costs are affected by slack---as labor cost is lower when the labor market is slacker. Hence, slack also appears in the Phillips curve, but only indirectly through marginal costs. See for instance \citet{KL07}, \citet{RW08}, \citet{TZ09}, \citet{T09}, \citet{BG10}, \citet{FMZ13}, and \citet{M14}.}

The economy is a matching economy composed of a measure one of large households, indexed by $j\in [0,1]$. Households purchase and consume services produced by workers from other households; they cannot consume the services of their own workers.\footnote{This is in the spirit of \citet{D82a}, who developed a model of the product market in which self-employed workers must trade because they cannot consume their own production. Modeling the US economy as a service economy is an accurate simplification: in 2024, 86\% of US workers were employed in service-providing industries \citep{USGOOD,SRVPRD}.} All trades are mediated by a matching function, which produces unemployment.

\subsection{Matching market for services}

In each household $j$, $h_j>0$ workers are part of the labor force; the aggregate labor force is $h = \int_0^1 h_j \,dj$. Each employed worker produces $k>0$ services per unit time, so the economy's productive capacity is $kh$ services per unit time.

Services are sold through long-term worker-household relationships. Once a worker has matched with a household, they become a full-time employee of the household, where they produce $k$ services per unit time. They remain so until they separate, which occurs at rate $\l>0$.

To recruit workers, each household $j$ posts $V_{j}$ help-wanted ads. The ads specify the price $P_j$ that the household plans to pay for the services purchased. Each help-wanted ad is serviced by $\k>0$ recruiters per unit time; the recruiters are employees of household $j$ devoted to recruiting other workers.

Job search is directed: workers choose which household to apply to. In that way, each household $j$ constitutes a submarket. Each jobseeker decides which submarket to join based on the price $P_j$ offered for services and the submarket tightness $\t_j = V_j/U_j$, where $U_j$ is the number of unemployed workers vying for jobs in household $j$. The tightness matters because it determines how quickly workers find a job. Jobseekers therefore trade off how much they are paid on the job with how long it takes to find the job.

In each submarket $j$, a matching function $m(U_j,V_j)$ determines the flow of new matches from the number of jobseekers and help-wanted ads. The matching function has constant returns to scale and is increasing in both arguments \citep{PP01}. Submarket tightness determines the job-finding rate, $f(\t_j) = m(1,\t_j)$, which is increasing in $\t_j$, and the ad-filling rate, $q(\t_j) = m(1/\t_j,1) = f(\t_j)/\t_j$, which is decreasing in $\t_j$. We denote by $\h(\t)\in(0,1)$ the elasticity of the matching function with respect to unemployment, which we assume to be increasing in tightness.\footnote{The Cobb-Douglas, constant-elasticity-of-substitution, and urn-ball matching functions all have this property \citep[chapter~3]{M26}.} Accordingly, the elasticity of $f(\t)$ is $1-\h(\t)$ and the elasticity of $q(\t)$ is $-\h(\t)$.

\subsection{Balanced flows: unemployment, vacancies, and matching wedge}\label{s:flows}

The unemployment rate in submarket $j$ is $u_j = U_j/h_j$, where $h_j$ is the labor force attached to submarket $j$. We assume that labor flows in each submarket are balanced: the number of workers who find a job with household $j$ at any point in time, $f(\t_j) U_j$, equals the number who leave their jobs, $\l (1-u_j) h_j$. Balanced flows imply that the unemployment rate is a function of submarket tightness, $u_j = u(\t_j)$, with
\begin{equation}
u(\t) = \frac{\l}{\l + f(\t)}.
\label{e:u}\end{equation}
The unemployment rate is strictly decreasing in tightness: when the market is tighter, workers find jobs faster, so fewer workers are unemployed at any point in time. The ad rate in submarket $j$ is $v_j = V_j/h_j$; with balanced flows it is given by $v(\t) = \t u(\t)$. As tightness moves, the pair $[u(\t), v(\t)]$ traces out the Beveridge curve of the model.

Balanced flows also determine how much labor is devoted to recruiting. Household $j$'s $l_j$ employees produce $y_j = k l_j$ services, but the $V_j$ help-wanted ads absorb $\k$ recruiters each, so a total of $k \k V_j$ services are devoted to recruiting instead of consumption. Household $j$'s consumption is therefore $c_j = k \bs{l_j - \k V_j}$, which can be written
\begin{equation}
c_j = k h_j \bs{1 - u_j - \k v_j}.
\label{e:cj}\end{equation}
To maintain $l_j$ employees, the household must fill $\l l_j$ ads per unit time, which requires $V_j = \l l_j / q(\t_j)$ ads. Hence output and consumption are related by a matching wedge:
\begin{equation}
y_{j} = [1+\tau(\t_j)] c_{j},
\label{e:yj}\end{equation}
where the matching wedge is
\begin{equation}
\tau(\t) = \frac{\k\l}{q(\t)-\k\l}.
\label{e:tau}\end{equation}

The matching wedge is positive and strictly increasing in tightness: in a tighter market, ads take longer to fill, so more labor must be devoted to recruiting for each unit of consumption. The wedge diverges when tightness reaches the upper bound $\bar{\t}$, defined by $q(\bar{\t}) = \k\l$; at that point all employed labor is devoted to recruiting. Combining \eqref{e:cj} and \eqref{e:tau}, consumption satisfies
\begin{equation}
c_j = \frac{1-u(\t_j)}{1+\tau(\t_j)} k h_j,\qquad \frac{1-u(\t_j)}{1+\tau(\t_j)} = 1 - u(\t_j) - \k v(\t_j). 
\label{e:c}\end{equation}

Two elasticities will be used repeatedly. First, from \eqref{e:u}, the elasticity of the unemployment rate with respect to tightness is
\begin{equation}
\e^u_{\t} = - \frac{f(\t)}{\l+f(\t)}\bs{1-\h(\t)} = - \bs{1-\h(\t)} \bs{1- u(\t)},
\label{e:uelasticity}\end{equation}
which also implies $\e^{1-u}_{\t} = \bs{1-\h(\t)} u(\t)$. Second, from \eqref{e:tau}, $1 + \tau = q/(q - \k\l)$, so the elasticity of $1+\tau(\t)$ with respect to tightness is
\begin{equation}
\e^{1+\tau}_{\t} = \e^{q}_{\t}\bs{1 - (1+\tau)} = \h(\t) \tau(\t).
\label{e:tauelasticity}\end{equation}

\subsection{Efficient allocation}\label{s:efficiency}

Consider a social planner who chooses tightness to maximize consumption of services---the only source of welfare in this economy.\footnote{Unemployed workers produce nothing in the model. This is consistent with the evidence: the value of jobseekers' home production, net of the psychological cost of idleness, is negligible \citep[pp.~333--334]{MS24}.} From \eqref{e:c}, aggregate consumption is $c(\t) = kh\bs{1-u(\t)}/\bs{1+\tau(\t)}$, so
\begin{equation*}
\e^{c}_{\t} = \e^{1-u}_{\t} - \e^{1+\tau}_{\t} = \bs{1-\h(\t)} u(\t) - \h(\t) \tau(\t).
\end{equation*}
The elasticity is positive at low tightness and negative at high tightness, so consumption is maximized at the efficient tightness $\t^*$, where the two terms offset:
\begin{equation}
\frac{\tau(\t^*)}{u(\t^*)} = \frac{1-\h(\t^*)}{\h(\t^*)}.
\label{e:efficiency}\end{equation}
The efficiency condition balances the two costs of slack: at $\t^*$, the labor lost to unemployment and the labor lost to recruiting are optimally traded off. When the economy is inefficiently slack ($\t<\t^*$), the wedge-to-unemployment ratio $\tau/u$ is below its efficient level $(1-\h)/\h$; when the economy is inefficiently tight ($\t>\t^*$), the ratio is above it.

The efficiency condition \eqref{e:efficiency} is the model counterpart of the sufficient-statistic formula \eqref{e:thetastar}. Indeed, using \eqref{e:u} and \eqref{e:tau}, condition \eqref{e:efficiency} can be rearranged as $\k \t \bs{\l + \h(\t) f(\t)} = \bs{1-\h(\t)}f(\t)$, which is $\t^*\b(\t^*)\k = 1$, where
\begin{equation}
\b(\t) = \frac{1}{1-\h(\t)} \bs{\h(\t)+\frac{u(\t)}{1-u(\t)}}
\label{e:beta}\end{equation}
is the elasticity of the model's Beveridge curve. In the model, the Beveridge curve is not exactly a rectangular hyperbola---its elasticity varies slightly with tightness---so the efficient tightness is not exactly 1.\footnote{In the calibrated model with a standard Cobb-Douglas matching function, $\b(\t^*) = 0.98$ and $\t^* = 1.03$. The difference is immaterial.} The FERU of the model is the unemployment rate at the efficient tightness: $u^* = u(\t^*)$.

\subsection{Directed search and the price-tightness tradeoff}\label{s:directed}

Households observe the prices and tightnesses in all submarkets and send their jobseekers to the submarkets that suit them best. This arbitrage links prices and tightnesses across submarkets---and this link is the basis for the Phillips curve.

When workers are unemployed, they receive no income; when they are employed, they produce $k$ services per unit time, each paid a price $P_j$. In submarket $j$, a fraction $1-u(\t_j)$ of the attached workers are employed at any point in time, so the expected income per worker in submarket $j$ is $k P_j \bs{1-u(\t_j)}$. Households send their workers to the submarkets offering the highest expected income per worker. In their role as employers, households are aware of this arbitrage, so they offer prices that allow them to compete with other households. Arbitrage therefore imposes that $P_j \bs{1-u(\t_j)}$ is the same across all submarkets:
\begin{equation}
P_j \bs{1 - u(\t_j)} = P \bs{1-u(\t)},
\label{e:arbitrage}\end{equation}
where $P$ is the price level and $\t$ is aggregate tightness---the ratio of the aggregate number of help-wanted ads to the aggregate number of jobseekers. If a household offered a lower expected income than the rest, it would simply attract no applicants.

The arbitrage condition \eqref{e:arbitrage} determines the tightness faced by a household as a function of the relative price it posts, $p_j = P_j/P$:
\begin{equation}
\t_j(p_j) = u^{-1}\of{1-\frac{1-u(\t)}{p_j}}.
\label{e:demand}\end{equation}
Since $u^{-1}$ is strictly decreasing, submarket tightness is decreasing in the relative price: a household that posts a generous price attracts many jobseekers, so it faces a slack submarket and recruits easily; a household that posts a stingy price attracts few jobseekers, so it faces a tight submarket and recruits slowly. In that way, households have monopsony power, and they face a downward-sloping demand-like curve in the price-tightness plane. Taking logs in \eqref{e:arbitrage} and differentiating, $1 + \e^{1-u}_{\t}\cdot \e^{\t}_{p} = 0$, so the elasticity of the curve is
\begin{equation}
\e^{\t}_{p} = \frac{-1}{[1-\h(\t_j)] u(\t_j)},
\label{e:demandelasticity}\end{equation}
where we used $\e^{1-u}_{\t} = (1-\h) u$ from section~\ref{s:flows}.

The tradeoff faced by household $j$ is then clear. All the services it purchases are paid the price $P_j$, so its expenditure is $P_j y_j = [1+\tau(\t_j)] P_j c_j$. The effective price of consumed services is not $P_j$ but $[1+\tau(\t_j)] P_j$: it includes the cost of the services consumed and the cost of the services devoted to recruiting. A low posted price $P_j$ economizes on the price per service but imposes a high tightness $\t_j$ and thus a high matching wedge $\tau(\t_j)$.

\subsection{Efficiency without price-adjustment costs}\label{s:m97}

As a benchmark, consider a household that is free to set any price. It minimizes the effective price of services, $[1+\tau(\t_j)] P_j$, subject to the arbitrage condition \eqref{e:arbitrage}. Using the arbitrage condition, the effective price can be written
\begin{equation*}
[1+\tau(\t_j)] P_j = P \bs{1-u(\t)} \cdot \frac{1+\tau(\t_j)}{1-u(\t_j)}.
\end{equation*}
Hence, by choosing its price, the household effectively chooses submarket tightness $\t_j$ to maximize $\bs{1-u(\t_j)}/[1+\tau(\t_j)]$---which, by \eqref{e:c}, is the same as maximizing $1 - u_j - \k v_j$. This is the social planner's problem of section~\ref{s:efficiency}, so the household selects the efficient tightness: $\t_j = \t^*$, where $\t^*$ satisfies \eqref{e:efficiency}. Here we have just recovered the efficiency result of \citet{M97}: directed search produces efficient outcomes, because pricing households internalize the planner's tradeoff between unemployment and recruiting.

\subsection{Price rigidity}\label{s:rigidity}

Generally, however, the labor market is not efficient: as figure~\ref{f:gap} shows, the US unemployment gap is large and countercyclical. To generate inefficient fluctuations, prices must be somewhat rigid.

Because search is directed, prices are chosen optimally by buyers---they are not set by norms---so we introduce rigidity through a cost of changing prices. As in \citet{R82}, household $j$ incurs a quadratic cost when inflation in its submarket, $\pi_{j}(t) = \dot{P}_{j}(t)/P_{j}(t)$, departs from normal inflation $\pi^n$---the inflation that people expect and consider fair. The flow disutility is
\begin{equation*}
\frac{\vp}{2} \bp{\pi_j-\pi^n}^2,
\end{equation*}
where the parameter $\vp>0$ governs the price-adjustment cost.

What does this cost capture? When prices grow less than normal, the workers of household $j$ feel shortchanged: the price $P_j$ directly determines their pay, $P_j k$, and workers' morale---and performance---dips when their pay does not grow as expected \citep{B05}. The quadratic cost reflects the cost of restoring morale, or the cost of diminished performance.\footnote{The typical Keynesian wage-floor idea is that workers do not tolerate a drop in the level of nominal wages. But as \citet[p.~93]{O81} notes, the same idea might apply to the growth rate of nominal wages, as we assume here: ``Workers are likely to develop some notion of how fast they expect their wages to rise.\dots If the employer stresses the norm, a Keynesian wage-floor phenomenon may apply to the rate of increase of wages, and not just to their level.''} Conversely, when prices grow more than normal, household $j$ is unhappy to pay more for the same services: higher-than-normal inflation upsets customers, who feel unfairly treated \citep{S97}. The quadratic cost captures this displeasure.\footnote{\citet{EMM21} provide a complete model of why price increases anger customers, and how sellers internalize such anger.}

\section{Households' problem}\label{s:households}

We now present and solve the optimization problem of households. Households have two decisions to make: how much to save, and how to set the prices of the services they purchase. From the optimal saving decision we will derive the aggregate demand curve; from the optimal pricing decision, the Phillips curve.

\subsection{Utility function}

Household $j$ cares about its consumption of services and its relative wealth, and it incurs the price-adjustment cost. The household's nominal wealth $B_j(t)$ is held in government bonds; its relative real wealth is $[B_j(t) - B(t)]/P(t)$, where $B(t) = \int_{0}^{1} B_{i}(t)\,di$ is average nominal wealth. Household $j$ maximizes the discounted sum of flow utilities,
\begin{equation}
\int_{0}^{\infty}\exp{-\d t} \bs{2 a \cdot \sqrt{c_j(t)} + \frac{B_j(t) - B(t)}{P(t)} - \frac{\vp}{2} \bp{\pi_j(t)-\pi^n}^2}\,dt,
\label{e:utility}\end{equation}
where $\d>0$ is the time discount rate and $a>0$ indicates the taste for consumption relative to the other components of utility.\footnote{The results do not rely on the specific functional form chosen here for the utility function. For instance, the results carry over with a more general utility function over consumption of the form $a \cdot c_j(t)^{1-\a}/(1-\a)$ with $\a\in(0,1)$, as in \citet{MS22}. The results also carry over if households value absolute real wealth $B_j(t)/P(t)$ instead of relative real wealth, as in \citet{M18}.}

The unusual ingredient is the presence of relative wealth in the utility function. People save not only to smooth consumption but also because they enjoy the social status conferred by wealth---an observation running from \citet{K19}, \citet{F30}, and \citet{L55} to modern neuroscientific evidence \citep{CLP05}. In fact, in a recent survey of 2,000 older Americans analyzed by \citet{C21}, more than half of respondents do not plan to spend down their wealth in retirement, and about 65\% agree that ``saving as much as I can makes me happy and fulfilled.'' Beyond realism, wealth in the utility function makes aggregate demand nondegenerate: it allows the model to accommodate interest-rate pegs, permanent zero-lower-bound episodes, and other policies that would produce anomalies in the New Keynesian model \citep{MS21a}.

\subsection{Budget constraint}

Household $j$'s nominal budget constraint is
\begin{equation*}
\dot{B}_{j}(t) = i(t) B_{j}(t) + \int_0^1 P_{i}(t) y_{ij}(t)\,di - P_{j}(t) y_{j}(t) - T(t),
\end{equation*}
where $i(t)$ is the nominal interest rate on bonds, $y_{ij}$ is the quantity of services sold by household $j$'s workers in submarket $i$, $y_j$ is the quantity of services purchased by household $j$, and $T(t)$ is a lump-sum tax that finances the interest payments on government bonds.

The two flow terms simplify. Since a fraction $1-u(\t_i)$ of the workers sent to submarket $i$ are employed, and the arbitrage condition \eqref{e:arbitrage} equalizes expected income across submarkets, the household's labor income is $\int_0^1 P_{i} y_{ij}\,di = P \bs{1-u(\t)} k h_j$. And given the matching wedge \eqref{e:tau} and the price-tightness link \eqref{e:demand}, the household's expenditure is $P_{j} y_{j} = P_{j} [1+\tau(\t_j(P_j/P))] c_j$. Translating the budget constraint into real terms, with real wealth $b_j = B_j/P$ and real interest rate $r(t) = i(t) - \pi(t)$, where $\pi = \dot{P}/P$ is aggregate inflation:
\begin{equation}
\dot{b}_{j}(t) = r(t) b_{j}(t) + \bs{1-u(\t(t))} k h_j - p_{j}(t) [1+\tau(\t_j(p_j(t)))] c_j(t) - \frac{T(t)}{P(t)}.
\label{e:budget}\end{equation}

Besides, household $j$ faces the law of motion of its relative price $p_j = P_j/P$:
\begin{equation}
\dot{p}_j(t) = \bs{\pi_j(t) - \pi(t)} p_j(t).
\label{e:pricelaw}\end{equation}

\subsection{First-order conditions}

Household $j$ chooses paths for consumption $c_j(t)$ and inflation $\pi_j(t)$ to maximize \eqref{e:utility} subject to \eqref{e:budget} and \eqref{e:pricelaw}. We solve this optimal-control problem with the Hamiltonian
\begin{align*}
\Hc &= 2a \cdot \sqrt{c_j}+ (b_j-b) - \frac{\vp}{2} (\pi_j-\pi^n)^2 \\
& +\p^b_j \bc{r b_{j} + \bs{1-u(\t)} k h_j -\bs{1+\tau(\t_j(p_j))} p_{j} c_j - \frac{T}{P}} +\p^p_j (\pi_j - \pi) p_j,
\end{align*}
where the control variables are consumption $c_j$ and inflation $\pi_j$; the state variables are real wealth $b_j$ and relative price $p_j$; and the costate variables are $\p^b_j$ and $\p^p_j$.

The first-order condition with respect to consumption, $\pdx{\Hc}{c_j}=0$, gives
\begin{equation}
\frac{a}{\sqrt{c_j}} = \p^b_j \bs{1 + \tau(\t_j)} p_j.
\label{e:foc1}\end{equation}
The first-order condition with respect to inflation, $\pdx{\Hc}{\pi_j}=0$, gives
\begin{equation}
\vp (\pi_j-\pi^n) = \p^p_j p_j.
\label{e:foc2}\end{equation}
The first-order condition with respect to real wealth, $\pdx{\Hc}{b_j}=\d\p^b_j -\dot{\p}^b_j$, gives
\begin{equation}
\dot{\p}^b_j = (\d - r) \p^b_j - 1.
\label{e:foc3}\end{equation}
The final first-order condition, with respect to the relative price, is $\pdx{\Hc}{p_j}=\d\p^p_j -\dot{\p}^p_j$. Expanding the derivative of the expenditure term, the key object is
\begin{equation*}
p_j \frac{\tau'(\t_j)}{1+\tau(\t_j)} \t_j'(p_j) = \e^{1+\tau}_{\t} \cdot \e^{\t}_{p} = -\frac{\h(\t_j)}{1-\h(\t_j)} \cdot \frac{\tau(\t_j)}{u(\t_j)},
\end{equation*}
where we combined the elasticities \eqref{e:tauelasticity} and \eqref{e:demandelasticity}. Noting also that $[1+\tau(\t_j)]c_j = y_j$, the condition simplifies to
\begin{equation}
\dot{\p}^p_j = (\d+\pi- \pi_j) \p^p_j + \p^b_j y_j \bs{1 -\frac{\h(\t_j)}{1-\h(\t_j)} \cdot \frac{\tau(\t_j)}{u(\t_j)}}.
\label{e:foc4}\end{equation}
The term in square brackets is central to what follows: by the efficiency condition \eqref{e:efficiency}, it is zero when submarket tightness is efficient, positive when the submarket is inefficiently slack, and negative when it is inefficiently tight.

\subsection{Symmetric solution}

We focus on a symmetric solution, in which all households behave the same and all submarkets are the same: all submarket tightnesses equal aggregate tightness $\t$, all relative prices equal 1, and all submarket inflation rates equal aggregate inflation $\pi$. Dropping the index $j$, the first-order conditions become
\begin{align}
\frac{a}{\sqrt{c(t)}} &= \p^b(t) \bs{1 + \tau(\t(t))}\label{e:foc1s}\\
\vp \bs{\pi(t)-\pi^n} &= \p^p(t)\label{e:foc2s}\\
\dot{\p}^b(t) &= \bs{\d - r(t)} \p^b(t) - 1\label{e:foc3s}\\
\dot{\p}^p(t) &= \d \p^p(t) + \p^b(t) y(t) \bs{1 -\frac{\h(\t(t))}{1-\h(\t(t))} \cdot \frac{\tau(\t(t))}{u(\t(t))}}.\label{e:foc4s}
\end{align}

\section{Solution of the model in normal times}\label{s:normal}

We now solve the model in normal times, when the nominal interest rate is positive, so the zero lower bound is not binding. Section~\ref{s:zlb} covers the zero lower bound.

\subsection{Monetary policy}

Monetary policy has two dimensions in the model: setting interest rates and communicating about future inflation.

First, the central bank maintains a fixed real interest rate $r<\d$, by setting the nominal interest rate to $i(t) = r + \pi(t)$ at every point in time. There is no Taylor rule: the central bank adjusts the nominal rate one-for-one with inflation to keep fixed the price that matters for aggregate demand---the real interest rate. The Taylor rule appears in the New Keynesian model mostly for convenience, to ensure the uniqueness of the solution \citep{C11}; here the solution is unique without it, so we dispense with it. 

For now we focus on normal times, when the nominal interest rate is positive, or the real rate is high enough: $r> -\pi(t)$. When the economy is too depressed for the nominal interest rate to be positive, the zero lower bound on nominal interest rates binds: $i(t) = 0$. We study this situation in section~\ref{s:zlb}.

Second, through communication and management of expectations, the central bank anchors normal inflation at its inflation target: $\pi^n = \pi^*$, with $\pi^* = 2\%$ in the United States \citep{FOMC26}. This assumption is central to the divine coincidence, and it is realistic: since 1995, the long-run inflation forecast elicited in the Survey of Professional Forecasters has been exceedingly stable, ever so slightly above 2\% \citep[figure~2]{HHN22}. Section~\ref{s:communication} explores what happens when the anchoring fails.

\subsection{Aggregate demand}

\begin{figure}[t]
\subcaptionbox{Phase line of the Euler equation \label{f:phaseline}}{\normalgraphic[7]{\npdf}}\hfill
\subcaptionbox{IS-MP representation \label{f:ismp}}{\normalgraphic[8]{\npdf}}
\caption{Construction of the aggregate demand curve}
\note{The Euler equation is given by \eqref{e:foc3s}; its equilibrium point is $1/(\d-r)$. The IS curve is given by \eqref{e:ad}.}
\label{f:ad}\end{figure}

Aggregate demand is determined by the two first-order conditions describing consumption and saving: \eqref{e:foc1s} and \eqref{e:foc3s}. Because the central bank fixes the real interest rate, this block is isolated from inflation: inflation matters for aggregate demand only through the real rate, which is fixed.

The differential equation \eqref{e:foc3s} is the Euler equation of the model. It is an autonomous linear differential equation in $\p^b$; since $\d-r>0$, it is a source, with equilibrium point
\begin{equation}
\p^b = \frac{1}{\d - r}.
\label{e:psib}\end{equation}
The phase line of the equation is plotted in figure~\ref{f:phaseline}. The costate variable $\p^b$ is not predetermined, so it can jump at any time. If it jumped above the equilibrium point, it would diverge upward, so consumption would fall to zero, violating feasibility; if it jumped below, consumption would diverge, eventually violating the resource constraint. Hence the costate jumps at time 0 to the equilibrium point \eqref{e:psib} and stays there.\footnote{The equilibrium value \eqref{e:psib} has a simple interpretation: households keep consumption constant only if the returns on wealth---financial returns of $r\p^b$ plus hedonic returns of $1$---offset time discounting $\d \p^b$.} 

Equation \eqref{e:foc1s} says that at any point in time consumption is a function of the costate variable:
\begin{equation*}
c(t) = \bs{\frac{a}{\p^b(t) \bs{1 + \tau(\t(t))}}}^2
\end{equation*}
Plugging \eqref{e:psib} into this equation gives consumption, and multiplying by $1+\tau(\t)$ gives the aggregate demand curve:
\begin{equation}
y^{d}(\t,r)= \frac{\bs{(\d-r) a}^{2}}{1 +\tau(\t)}.
\label{e:ad}\end{equation}
The aggregate demand curve gives the quantity of services demanded by households when they save optimally. It is decreasing in tightness---because a tighter economy imposes a larger matching wedge, making consumption effectively more expensive---and decreasing in the real interest rate---because a higher interest rate makes saving more attractive.

Figure \ref{f:ismp} shows that our model provides a microfoundation for the IS-MP model proposed by \citet{R00}---itself built from the IS-LM model invented by \citet{H37}.\footnote{The IS-MP model is a popular model to teach business cycles to undergraduate students. It is used for instance by \citet{M15}, \citet{J25}, or \citet{SW26}.} The IS curve is typically difficult to reconcile with utility-maximizing consumption-saving behavior \citep[p.~532]{BF89}. But it appears naturally once wealth enters the utility function. In \eqref{e:ad}, output is a nondegenerate, decreasing function of the real interest rate: an IS curve. The MP curve is a fixed real interest rate selected by the central bank. The intersection of the IS and MP curves give output as a function of tightness only, which is the aggregate demand curve $y^d(\t)$ plotted in figure~\ref{f:adas}.

\subsection{Aggregate supply}

The aggregate supply is the quantity of services provided when flows are balanced:
\begin{equation}
y^s(\t) = \bs{1-u(\t)} k h.
\label{e:as}\end{equation}
The aggregate supply is increasing in tightness---in a tighter economy, fewer workers are unemployed---and it is sharply convex in a tightness-output plane: flat when tightness is low, steep when tightness is high (figure~\ref{f:adas}).

\subsection{Aggregate tightness}

\begin{figure}[t]
\normalgraphic[9]{\npdf}
\caption{Aggregate demand, aggregate supply, and model solution in normal times}
\note{The aggregate supply curve is given by \eqref{e:as}. The aggregate demand curve is given by \eqref{e:ad}. The intersection of the two curves determines aggregate tightness and output; unemployment is read off the distance between output and capacity $kh$.}
\label{f:adas}\end{figure}

The aggregate tightness that solves the model equalizes aggregate demand and supply:
\begin{equation}
y^d(\t,r) = y^s(\t).
\label{e:solution}\end{equation}
Since demand is decreasing in tightness and supply is increasing, the solution is unique. It is illustrated in figure~\ref{f:adas}. Equation~\eqref{e:solution} implicitly defines tightness as a function of the real interest rate, $\t(r)$; from tightness, all quantities follow: output $y = y^s(\t)$, unemployment $u(\t)$, consumption $c(\t)$.

\subsection{Phillips curve}\label{s:phillips}

Inflation is determined by the two remaining first-order conditions, which describe optimal pricing: \eqref{e:foc2s} and \eqref{e:foc4s}.

Equation \eqref{e:foc2s} says that inflation is a linear function of the costate variable on the relative price: 
\begin{equation*}
\pi(t) = \pi^* + \frac{\p^p(t)}{\vp}. 
\end{equation*}
The costate variable follows the differential equation \eqref{e:foc4s}, which is linear and a source since $\d>0$. If the costate did not jump to the equilibrium point of the equation, it would explode, dragging inflation to infinity and generating infinitely negative utility. So the costate jumps to the equilibrium point:
\begin{equation*}
\p^p = \frac{y(\t)}{\d (\d-r)} \bs{\frac{\h(\t)}{1-\h(\t)} \cdot \frac{\tau(\t)}{u(\t)} - 1},
\end{equation*}
where we used $\p^b = 1/(\d-r)$. Accordingly, inflation jumps at time 0 to
\begin{equation}
\pi = \pi^* + \frac{y(\t)}{\vp \d (\d-r)} \bs{\frac{\h(\t)}{1-\h(\t)} \cdot \frac{\tau(\t)}{u(\t)} - 1}.
\label{e:pc}\end{equation}

Equation \eqref{e:pc} is the Phillips curve of the model. It links inflation to aggregate tightness, which is itself determined by aggregate demand and supply. The curve has three main properties.

First, the divine coincidence holds. The term in square brackets measures the productive inefficiency of the economy: by the efficiency condition \eqref{e:efficiency}, it is zero when tightness is efficient. So inflation is on target ($\pi = \pi^*$) if and only if the economy is at full employment ($\t = \t^*$, $u = u^*$). When the economy is inefficiently tight, the term is positive and inflation is above target; when the economy is inefficiently slack, the term is negative and inflation is below target. Thanks to the coincidence, the central bank faces no tradeoff between inflation and unemployment: by bringing unemployment to the FERU, it mechanically brings inflation to target. The coincidence arises because people take the inflation target as normal inflation; section~\ref{s:communication} shows how the coincidence breaks when they do not.

Second, the slope of the Phillips curve is governed by the price-adjustment cost $\vp$. When prices are very costly to change, $\vp$ is large and inflation responds little to departures from efficiency. When prices are easy to change, inflation responds forcefully. In the limit $\vp \to \infty$, inflation is fixed at $\pi^*$; in the limit $\vp \to 0$, the Phillips curve is vertical at $\t^*$ and the economy is always efficient, as in section~\ref{s:m97}.

Third, inflation is strictly increasing in tightness at full employment and above. Output $y(\t)$, the matching elasticity $\h(\t)$, and the matching wedge $\tau(\t)$ are increasing in tightness, while the unemployment rate $u(\t)$ is decreasing, so above full employment the bracket is positive and growing while output is positive and growing: inflation grows with tightness. Around full employment the bracket is zero, so the slope of the Phillips curve is determined by the derivative of the bracket, which is strictly positive. Below full employment, two forces oppose: as tightness falls, output falls while the (negative) bracket grows in magnitude. The growing inefficiency pulls inflation further below target; the falling output, which scales the bracket, pulls inflation back toward target. Near full employment the bracket dominates and inflation still increases with tightness; at very low tightness the output term can dominate and inflation may locally decrease with tightness.\footnote{In the calibrated model, this non-monotonicity only arises at tightness below the extremely low value of 0.08---so in practice inflation is increasing in tightness.}

What is the intuition for the Phillips curve? When inflation is above normal, a buyer could reduce its price-adjustment cost by lowering its rate of inflation. Since pricing is optimal, there cannot exist any profitable deviation, so lowering inflation must be costly for another reason: with lower inflation, the buyer's posted price would drop relative to other buyers' prices, and this price reduction must hurt. That is only possible if the price is already below the effective-price-minimizing level---which means, since submarket tightness is inversely related to the posted price, that tightness is above the efficient level (section~\ref{s:m97}). Hence, when inflation is above normal, the economy must be inefficiently tight. Symmetrically, when inflation is below normal, the economy must be inefficiently slack. Price dynamics are thus driven by slack: when workers are scarce, buyers must post higher prices to attract them; when workers are abundant, buyers can lower the prices they pay.\footnote{We have assumed that buyers set prices. The results are the same if sellers set the prices instead: workers would raise their prices when they are in high demand and lower them when demand slumps.}

\section{Quantitative Phillips curve}\label{s:quantitative}

To assess the shape of the Phillips curve, we calibrate the model to the US economy and we plot the curve in the tightness-inflation plane and the unemployment-inflation plane.

\subsection{Calibration}\label{s:calibration}

\begin{table}[t]
\caption{Calibrated values of the parameters and variables in the Phillips curve}
\begin{tabular*}{\textwidth}{@{\extracolsep{\fill}}llp{6cm}}\toprule
Value & Description & Source or target\\
\midrule
\multicolumn{3}{c}{A. Normalized parameters}\\
$k = 1$ & Labor productivity & Normalization\\
$h = 1$ & Labor force & Normalization\\
\midrule
\multicolumn{3}{c}{B. Parameters calibrated from direct measurement}\\
$\k = 1$ & Recruiting cost & \citet{MS24}\\
$\l = 3.6\%$ & Monthly job-separation rate & \citet{JTSTSR}\\
$\d = 43\%$ & Annual time discount rate & \citet{FLO02,AHL14}\\
$r = 0.5\%$ & Annual real interest rate & \citet{W17}\\
\midrule
\multicolumn{3}{c}{C. Parameters calibrated to match target}\\
$\s=0.47$ & Matching elasticity & Beveridge elasticity $=1$ (figure~\ref{f:hyperbola})\\
$\m = 0.75$ & Monthly matching efficacy & Unemployment rate $= 5.7\%$\\
$\vp = 149$ & Price-adjustment cost & Repricing cost $=1.22\%$ of revenue \citep{ZRL04}\\
\midrule
\multicolumn{3}{c}{D. Variables calibrated from historical evidence}\\
$\t = 0.65$ & Labor market tightness & Average over 1948--2019\\
$u = 5.7\%$ & Unemployment rate & Average over 1948--2019\\
\midrule
\multicolumn{3}{c}{E. Implied variables}\\
$v = 3.7\%$ & Vacancy rate & $v = \t u$\\
$f(\t) = 0.60$ & Monthly job-finding rate & $f(\t) = \m\t^{1-\s}$\\
$q(\t) = 0.92$ & Monthly ad-filling rate & $q(\t) = \m\t^{-\s}$\\
$\tau(\t) = 4.1\%$ & Matching wedge & Equation~\eqref{e:tau}\\
$\t^* = 1.03$ & Efficient tightness & Equation~\eqref{e:efficiency}\\
$u^* = 4.5\%$ & FERU & $u^* = u(\t^*)$\\
\bottomrule\end{tabular*}
\label{t:calibration}\end{table}

The calibration is summarized in table~\ref{t:calibration}. As much as possible, we use US evidence from the 1948--2019 period, during which the labor market operated in a consistent fashion; we exclude the disruptions caused by the Great Depression, World War 2, and the coronavirus pandemic.

Several parameters are directly measured. The recruiting cost is set to $\k = 1$ worker per vacancy, as measured by the recruiting surveys discussed in section~\ref{s:feru}. The job-separation rate is set to $\l = 3.6\%$ per month, the 2001--2019 average of the total-separation rate measured in JOLTS \citep{JTSTSR}. The real interest rate is set to $r = 0.5\%$ per year, the US natural rate estimated by \citet[table~1]{W17}. Following \citet[p.~213]{MS21b}, the discount rate is set to $\d = 43\%$ per year. This estimate is based on the annual discount rates estimated in the 54 different experimental studies surveyed by \citet[table~1]{FLO02} and \citet[table~3]{AHL14}.\footnote{With wealth in the utility function, the discount rate is not tied to the average real interest rate, so it can be calibrated from experimental evidence---which consistently points to double-digit discount rates \citep{CEL20}. Such impatience may not be surprising considering that annual interest rates on US credit cards are typically around 20\% \citep{R26}, with penalty rates near 30\% \citep{I26}.}

In addition, a few parameters are normalized. Labor productivity and labor force are normalized to $k = h = 1$ at the baseline. 

The remaining parameters are calibrated to match key US facts. The matching function is Cobb-Douglas, $m(U,V) = \m U^{\s}V^{1-\s}$, so the matching elasticity is $\h(\t) = \s$. We set $\s$ to generate a Beveridge elasticity of 1, as measured in figure~\ref{f:hyperbola}. Inverting \eqref{e:beta} at the average unemployment rate $u = 5.7\%$ gives $\s = 0.47$. We then set the matching efficacy $\m$ so that the unemployment rate is at its 1948--2019 average of 5.7\% when tightness is at its 1948--2019 average of 0.65; equation \eqref{e:u} gives $\m = 0.75$ per month. Under this calibration, the efficient tightness is $\t^* = 1.03$ and the FERU is $u^* = u(\t^*) = 4.5\%$.

The final parameter is the price-adjustment cost $\vp$, which determines the slope of the Phillips curve. We calibrate it from microevidence collected by \citet{ZRL04}. Using time-and-motion methods, they study the pricing process of a large industrial firm and find that the physical, managerial, and customer costs of changing prices amount to 1.22\% of the firm's revenue in a given year. That year, the firm changed the price of 25\% of its products, and most prices changed by about 4\%. Based on this evidence, we set $\vp$ so that if inflation is 4\% on 25\% of the services purchased by a household, the cost amounts to 1.22\% of the household's purchases in utility terms.

The marginal utility of consumption is $a/\sqrt{c}$. Through first-order condition \eqref{e:foc1} and the costate variable value \eqref{e:psib}, we know that the marginal utility equals $[1+\tau(\t)]/(\d-r)$. Hence we calibrate the price-adjustment cost such that:
\begin{equation*}
\frac{\vp}{2} \times 0.04^2 \times 0.25 = 0.0122 \times \frac{1+\tau(\t)}{\d-r}.
\end{equation*}
At the average tightness of $0.65$, $1+\tau(\t) = 1.041$. At the average real rate of $0.005$, $\d-r= 0.425$. Based on these values, we obtain $\vp = 149$.

\subsection{Inflation and tightness: divine coincidence}

\begin{figure}[t]
\subcaptionbox{Tightness gap versus inflation gap \label{f:pctightnessgap}}{\normalgraphic[1]{\npdf}}\hfill
\subcaptionbox{Unemployment gap versus inflation gap\label{f:pcunemploymentgap}}{\normalgraphic[2]{\npdf}}\fspace
\subcaptionbox{Tightness versus inflation \label{f:pctightness}}{\normalgraphic[3]{\npdf}}\hfill
\subcaptionbox{Unemployment versus inflation \label{f:pcunemployment}}{\normalgraphic[4]{\npdf}}
\caption{Phillips curves in the calibrated model}
\note{The Phillips curve is given by equation~\eqref{e:pc}. Tightness and unemployment are related by equation~\eqref{e:u}. The tightness gap is $\t - \t^*$; the unemployment gap is $u-u^*$; the inflation gap is $\pi - \pi^*$. The model is calibrated as in table~\ref{t:calibration}.}
\label{f:phillipscurves}\end{figure}

Figure~\ref{f:pctightnessgap} plots the Phillips curve \eqref{e:pc} in the tightness gap-inflation gap plane. The curve goes through the point of divine coincidence: the inflation gap is zero when the tightness gap is zero. The non-monotonicity discussed in section~\ref{s:phillips} is visible only at extremely low tightness (below 0.08). Otherwise the inflation gap is increasing in the tightness gap.

The Phillips curve is broadly linear in tightness over that range. For instance, when the tightness gap is $-0.5$, the inflation gap is $-1.6$pp, while when the tightness gap is $+0.5$, the inflation gap is $+1.7$pp. 

To illustrate the tightness and inflation levels produced by the model, and not just the gaps, figure~\ref{f:pctightness} plots the Phillips curve \eqref{e:pc} in the tightness-inflation plane. The Phillips curve goes through the point of divine coincidence, where tightness is $\t^*=1.03$ and inflation is $\pi^* = 2\%$. We see that inflation is increasing and almost linear in tightness over a wide range (as soon as tightness is above 0.08). 

\subsection{Inflation and unemployment: convexity}\label{s:convexity}

The picture changes once the Phillips curve is represented in traditional unemployment-inflation planes. Figure~\ref{f:pcunemploymentgap} plots the Phillips curve \eqref{e:pc} in the unemployment gap-inflation gap plane. The curve goes through the point of divine coincidence again: the inflation gap is zero when the unemployment gap is zero. And as expected, since unemployment and tightness are negatively related through \eqref{e:u}, the inflation gap is decreasing in the unemployment gap.

But inflation is now a strongly convex function of unemployment. From the point of divine coincidence, when the unemployment gap rises by 8pp, the inflation gap falls by less than 3pp. But when the unemployment gap drops by 2pp, the inflation gap rises by more than 6pp.\footnote{The shape of the simulated Phillips curve in figure~\ref{f:pcunemployment} is uncannily similar to the shape of the original curve estimated by \citet[figure~1]{P58} in the United Kingdom.} So the Phillips curve is much steeper when the economy is inefficiently tight than when it is inefficiently slack.

To illustrate the unemployment and inflation levels produced by the model, figure~\ref{f:pcunemployment} plots the Phillips curve \eqref{e:pc} in the unemployment-inflation plane. The curve crosses the point of divine coincidence, where inflation is $\pi^* = 2\%$ and unemployment is $u^* = 4.5\%$. The curve is much steeper when unemployment is inefficiently low than when it is inefficiently high. The inflation rate only falls to 0\% when the unemployment rate reaches 7\%, and then to -0.7\% when the unemployment rate climbs to 11\%. By contrast, the inflation rate rises to 4\% when the unemployment rate falls to 3.6\%, and spikes at 6.4\% when the unemployment rate further drops to 3\%.

The convexity is not an artifact of the calibration. It can be established analytically under the common assumption that the matching function is Cobb-Douglas and symmetric:
\begin{proposition}\label{r:convexity}
Assume that the matching function is Cobb-Douglas and symmetric: $m(u,v) = \m\sqrt{uv}$. Let $\bar{u} = \k\l^2/(\m^2+\k\l^2)$ be the unemployment rate at the upper tightness bound $\bar{\t}$. Then the Phillips curve $\pi(u)$ given by \eqref{e:pc} is strictly decreasing in $u$ around the FERU $u^*$ and strictly convex in $u$ on $(\bar{u},1)$.
\end{proposition}

The proof is in appendix~\ref{a:convexity}. The logic is simple. With a symmetric Cobb-Douglas matching function, the ratio of matching wedge to unemployment admits the closed form $\tau(u)/u = (1-\bar{u})/(u-\bar{u}) - 1/u$, which is strictly decreasing and strictly convex in $u$. The Phillips curve inherits this shape: inflation explodes as unemployment approaches its feasible minimum $\bar{u}$ and flattens out as unemployment grows.

The convexity means that the model produces relatively large fluctuations in inflation and small fluctuations in unemployment when the economy is inefficiently tight---and the reverse when the economy is inefficiently slack. This is the pattern visible in the post-pandemic data (figure~\ref{f:scatter}): the empirical Phillips curve drawn by the US economy in 2020--2024 is quite close to the Phillips curve simulated from the model (figure~\ref{f:pcunemploymentgap}). For the same fluctuations in the unemployment gap, the inflation gap varies by comparable amounts in the data and in the simulations. The convexity also explains why econometric studies detect a kink at the postwar FERU of 4.2\% (section~\ref{s:fact3}): a convex curve estimated with two linear branches yields a steep branch below the FERU and a flat branch above it.

\subsection{Nonlinearity of the model}

The Phillips curve is almost linear in tightness yet strongly convex in unemployment. How can the same curve have such different curvature in the two representations? The reconciliation is the Beveridge mapping $u(\t)$: unemployment is a decreasing and convex function of tightness, so an increasing and nearly linear $\pi(\t)$ generates a decreasing and convex $\pi(u)$.

The near-linearity of the Phillips curve in the tightness-inflation plane does not mean that the model is linear. The aggregate supply curve \eqref{e:as} is sharply nonlinear in the tightness-output plane: aggregate demand shocks have small effects on tightness in a slack economy---they affect mostly output---but large effects on tightness in a tight economy. Since inflation is linear in tightness, inflation behaves in the same way: it responds little to demand shocks when the economy is slack and much more when the economy is tight. Equivalently, in the unemployment representation: the supply curve is linear in unemployment, $y^s = (1-u)kh$, so unemployment responds similarly to demand shocks in booms and slumps, but inflation responds sharply to unemployment when the economy is tight---because the Phillips curve is convex---and barely when the economy is slack. Whether the nonlinearity is assigned to the aggregate supply curve or to the Phillips curve is a matter of representation; the economics is the same.

\section{Response to business-cycle shocks}\label{s:shocks}

We now use the model to study business-cycle shocks. Because all jump variables move instantly to their equilibrium values, the response to unexpected permanent shocks is obtained by comparative statics: the economy jumps from the old intersection of aggregate demand and supply to the new one, and inflation jumps along the Phillips curve.

\begin{figure}[t]
\subcaptionbox{Negative aggregate demand shock \label{f:adshock}}{\normalgraphic[10]{\npdf}}\hfill
\subcaptionbox{Negative aggregate supply shock \label{f:asshock}}{\normalgraphic[11]{\npdf}}
\caption{Response of the model to aggregate demand and supply shocks}
\note{The aggregate supply curve is given by equation~\eqref{e:as}. The aggregate demand curve is given by equation~\eqref{e:ad}. A negative aggregate demand shock is a decrease in the taste for consumption $a$; it rotates the aggregate demand curve inward. A negative aggregate supply shock is a decrease in productive capacity $kh$; it shifts the aggregate supply curve inward.}
\label{f:shocks}\end{figure}

\subsection{Aggregate demand shocks}

The most natural negative aggregate demand shock is a decrease in the taste for consumption $a$: households become thriftier, desiring to save more and consume less. The shock rotates the aggregate demand curve inward (figure~\ref{f:adshock}). Tightness and output fall; unemployment rises. Since the efficient tightness \eqref{e:efficiency} does not depend on $a$, the FERU is unchanged, so the unemployment gap widens one-for-one with unemployment. Along the Phillips curve, the drop in tightness lowers inflation: the economy slides down the curve, producing the standard negative comovement of unemployment and inflation.

The effects of demand shocks are sharply state-dependent. When the economy is slack, the aggregate supply curve is flat, so demand shocks move mostly output and unemployment, and barely inflation. When the economy is tight, the supply curve is steep, so demand shocks move mostly tightness and inflation, and barely unemployment. In the unemployment-inflation plane, this is just the convexity of the Phillips curve at work: demand shocks generate large unemployment movements and small inflation movements in slumps, and the reverse in booms. Both phases of the pandemic episode fit this pattern: the collapse of demand in 2020, hitting a slack economy, moved unemployment enormously and inflation little; the surge of demand in 2021--2022, hitting a tight economy, moved inflation enormously and unemployment little (figure~\ref{f:scatter}).

\subsection{Aggregate supply shocks}

A negative aggregate supply shock is a decrease in productive capacity $kh$---caused by a decrease in labor productivity $k$ or in the labor force $h$. The shock shifts the aggregate supply curve inward (figure~\ref{f:asshock}). Output falls but tightness rises: people want to consume a similar amount of services while fewer services are offered, so the economy gets tighter and unemployment falls. Along the Phillips curve, the rise in tightness raises inflation. Supply shocks are therefore stagflationary in the model: they lower output while raising inflation. And because the efficient tightness does not depend on $k$ or $h$, the drop in unemployment translates into a negative unemployment gap: the economy becomes inefficiently tight, which is why inflation rises.

\subsection{Beveridge shocks}\label{s:beveridge}

The third type of shock is specific to the Beveridgean model: shocks that shift the Beveridge curve. In the model, the position of the Beveridge curve is determined by the job-separation rate $\l$ and the matching efficacy $\m$. An increase in $\l$ or a decrease in $\m$ shifts the Beveridge curve outward: at any tightness, unemployment is higher.

A Beveridge shock differs from demand and supply shocks in one respect: it moves the FERU. The efficiency condition \eqref{e:efficiency} depends on $\l$ and $\m$ through $u(\t)$ and $\tau(\t)$, so an outward shift of the Beveridge curve raises the efficient unemployment rate $u^* = u(\t^*)$---as it did dramatically during the pandemic, when the FERU jumped from 4.0\% to 6.7\% in one quarter (figure~\ref{f:feru}). Because the point of divine coincidence sits at $(u^*, \pi^*)$, the unemployment-inflation Phillips curve shifts outward with the Beveridge curve: the same inflation now obtains at higher unemployment.

Beveridge shocks therefore provide a microfoundation for Phillips shocks---akin to the cost-push shocks of the New Keynesian literature. Here, the Phillips curve shifts whenever the Beveridge curve shifts, and shifts of the Beveridge curve are observable in the data: the Phillips-curve shifts are measurable in real time through the FERU formula $u^* = \sqrt{uv}$.

Critically, the divine coincidence survives Beveridge shocks: the coincidence point moves, but inflation remains on target if and only if unemployment is at the (new) FERU. No shock in the model can generate inflation at full employment. The tightness formulation of the Phillips curve is also far more stable than the unemployment formulation. The efficient tightness is pinned down by the Beveridge elasticity and the recruiting cost, $\t^* = 1/(\k\b)$, and both statistics are stable in US data. Thus, $\t^*$ barely moves when the Beveridge curve shifts, while $u^*$ moves a lot. This explains our second fact: why tightness predicts inflation so much better than unemployment, especially in the pandemic period when the Beveridge curve was unstable.

\section{Solution of the model at the zero lower bound}\label{s:zlb}

In normal times, the central bank sets the real interest rate, which separates the dynamics of tightness from those of inflation: tightness is determined by aggregate demand and supply, and inflation follows from the Phillips curve. At the zero lower bound (ZLB), this separation disappears. The nominal rate is stuck at 0, so the real rate is just negative inflation, $r = -\pi$, and tightness and inflation are jointly determined by a two-dimensional dynamical system.

\subsection{When does the ZLB bind?}

The economy reaches the ZLB after a large drop in aggregate demand---for instance a large decrease in the taste for consumption $a$. The ZLB binds when, even at a zero nominal rate, aggregate demand falls short of the efficient level of output: $y^d(\t^*,-\pi^*)<y^s(\t^*)$ (figure~\ref{f:zlbad}). This occurs when the taste for consumption satisfies
\begin{equation}
a < \frac{\sqrt{(1 + \tau^*) y^*}}{\d+\pi^*},
\label{e:zlbcondition}\end{equation}
where $y^*$ and $\tau^*$ are the efficient values of output and the matching wedge.

\subsection{Tightness-inflation dynamics}

\begin{figure}[t]
\subcaptionbox{Source at the zero lower bound \label{f:zlbhigh}}{\normalgraphic[5]{\npdf}}\hfill
\subcaptionbox{Source with lower aggregate demand \label{f:zlblow}}{\normalgraphic[6]{\npdf}}
\caption{Phase diagrams at the zero lower bound in the calibrated model}
\note{The $\dot{\t}=0$ nullcline is given by equation~\eqref{e:theta0}. The $\dot{\pi}=0$ nullcline is given by equation~\eqref{e:pi0}. The model is calibrated as in table~\ref{t:calibration}. In panel A, the taste for consumption is $a = 1.66$. In panel B, the taste for consumption is lower, set to $a = 1.62$.}
\label{f:zlbphase}\end{figure}

The dynamical system describing the model at the ZLB is derived in appendix~\ref{a:zlb}; here we report the main results. The system comprises two differential equations. The $\dot{\t}$ equation arises from the consumption-saving condition \eqref{e:foc3s}:
\begin{equation}
\frac{1}{2}\cdot\frac{\dot{\t}}{\t} [[1-\h(\t)] u(\t) + \h(\t) \tau(\t)] = \frac{\sqrt{[1+ \tau(\t)] y(\t)}}{a} - \d - \pi.
\label{e:thetadot}\end{equation}
The $\dot{\pi}$ equation arises from the pricing condition \eqref{e:foc4s}:
\begin{equation}
\dot{\pi} = \d \bs{\pi-\pi^*} + \frac{a}{\vp} \sqrt{c(\t)} \bs{1 -\frac{\h(\t)}{1-\h(\t)} \cdot \frac{\tau(\t)}{u(\t)}}.
\label{e:pidot}\end{equation}

Figure~\ref{f:zlbphase} displays the phase diagram of the system in the calibrated model. The $\dot{\t}=0$ nullcline combines the aggregate demand and supply curves: as inflation rises, the real rate falls, so aggregate demand moves up along aggregate supply, raising tightness. The nullcline is
\begin{equation}
\pi = \frac{1}{a} \sqrt{[1 + \tau(\t)] y(\t)} - \d.
\label{e:theta0}\end{equation}
The $\dot{\pi}=0$ nullcline is based on the Phillips curve, describing inflation as an increasing function of tightness over the relevant range:
\begin{equation}
\pi = \pi^* + \frac{a}{\d\vp} \sqrt{c(\t)} \bs{\frac{\h(\t)}{1-\h(\t)} \cdot \frac{\tau(\t)}{u(\t)} - 1}.
\label{e:pi0}\end{equation}
This nullcline remains below the inflation target on $(0,\t^*)$, and it is flatter---closer to $\pi = \pi^*$---when prices are more rigid (higher $\vp$).

The system always admits an equilibrium $(\t^z, \pi^z)$ with $\t^z \in (0,\t^*)$, and when price rigidity $\vp$ is large enough the equilibrium is unique (appendix~\ref{a:zlb}). Around the equilibrium, the system is a source: in all four quadrants of the phase diagram, trajectories move away from the equilibrium (figure~\ref{f:zlbphase}). The solution of the model at the ZLB is therefore unique, although monetary policy is completely passive: tightness and inflation must jump to the equilibrium and stay there; had they jumped anywhere else, they would have diverged.

\subsection{The ZLB is not topsy-turvy}

The uniqueness result means that the model can be used to study ZLB episodes of any duration---even permanent ones. The unique ZLB solution is a normal slump: tightness is below its efficient level, unemployment is above the FERU, and inflation is below target. Comparative statics remain normal as well. A negative aggregate demand shock---a further decrease in the taste for consumption $a$---shifts the $\dot{\t}=0$ nullcline inward (from figure~\ref{f:zlbhigh} to figure~\ref{f:zlblow}): tightness falls, unemployment rises, and inflation falls. Demand shocks operate at the ZLB just as they do in normal times.

This behavior is entirely different from the New Keynesian model, in which the ZLB is ``a topsy-turvy world, in which many of the usual rules of macroeconomics are stood on their head'' \citep[p.~1472]{EK12}. Indeed, when the ZLB lasts long, output and inflation collapse to implausible levels, the model admits no solution with a permanent ZLB, and policies acquire extraordinarily large effects \citep{MS21a}. None of these anomalies arise here. The difference comes from wealth in the utility function, which keeps aggregate demand well defined even under an interest-rate peg---and the ZLB is nothing but a peg at zero.

\section{Implications for monetary policy}\label{s:policy}

This section reviews the implications of the Beveridgean Phillips curve for monetary policy.

\subsection{Satisfying the dual mandate: the FERI}

\begin{figure}[t]
\subcaptionbox{Interest rate cuts \label{f:ratecut}}{\normalgraphic[12]{\npdf}}\hfill
\subcaptionbox{Aggregate demand at the ZLB \label{f:zlbad}}{\normalgraphic[13]{\npdf}}
\caption{Monetary policy in the model}
\note{The aggregate demand and supply curves are constructed in figure~\ref{f:adas}. In panel A, the gray cone indicates all the positions that the aggregate demand curve can reach when the nominal interest rate is reduced from its current level; the ZLB curve is the most outward position that the aggregate demand curve can reach, obtained when $i = 0$. In panel B, aggregate demand is so depressed that even at the zero lower bound it intersects aggregate supply below the efficient tightness.}
\label{f:monetary}\end{figure}

In the United States, the Federal Reserve has a dual mandate: maintaining full employment and price stability. In the model, both mandates are achieved at once. By selecting the real interest rate $r$, monetary policy can stimulate or depress aggregate demand (figure~\ref{f:ismp}). This policy move then shifts the aggregate demand curve along the aggregate supply curve, which changes aggregate tightness (figure~\ref{f:ratecut}). To satisfy the dual mandate, policy must bring the aggregate demand curve to the point where tightness is efficient. There, the Beveridge curve guarantees full employment ($u = u^*$) and the Phillips curve guarantees price stability ($\pi = \pi^*$).

We call the real interest rate that brings the economy to full employment the full-employment rate of interest (FERI), denoted $r^*$. Because of the divine coincidence, the FERI also delivers price stability. It solves $y^d(\t^*,r^*) = y^s(\t^*)$, which from \eqref{e:ad} and \eqref{e:as} gives
\begin{equation}
r^* = \d - \frac{1}{a}\cdot \sqrt{[1+\tau(\t^*)] y(\t^*)},
\label{e:rstar}\end{equation}
where $y(\t^*)$ is the efficient level of output. The corresponding nominal interest rate is $i^* = r^* + \pi^*$. The Fed can achieve its dual mandate whenever $i^*\ge 0$. If aggregate demand is so depressed---the taste for consumption $a$ so low---that $r^* < -\pi^*$, the dual-mandate policy would violate the zero lower bound; the central bank then sets $i = 0$, and the analysis of section~\ref{s:zlb} applies.

\subsection{Soft landing}\label{s:softlanding}

After the pandemic, inflation rose well above target---the inflation gap peaked at 4.9pp in 2022---while the labor market became inefficiently tight (figure~\ref{f:pandemic}). Once inflation was high, the policy debate centered on a mechanical question: could the Federal Reserve achieve a soft landing? That is, could the Fed bring inflation back down without a material rise in unemployment \citep{B23}?

In the model, inflation is governed by tightness through the Phillips curve, so cooling inflation requires reducing tightness. Monetary policy operates through aggregate demand: raising the real interest rate depresses demand, which reduces tightness along the aggregate supply curve. Since unemployment moves inversely with tightness, a strict soft landing is impossible: unemployment always rises when tightness falls. But the rise can be modest when the labor market starts from a very tight position, because the aggregate supply curve is convex: when tightness is high, a given reduction in tightness raises unemployment only a little.

The Beveridge curve makes the required rise in unemployment easy to quantify. By definition, $\t = v(u)/u$ along the Beveridge curve, so differentiating in elasticity form gives $1 = -\b \e^u_{\t} - \e^u_{\t}$, and therefore
\begin{equation*}
\e^u_{\t} = \frac{-1}{1+\b}.
\end{equation*}
In the United States, $\b=1$, so the elasticity of unemployment with respect to tightness is $-1/2$. Because the Beveridge curve is isoelastic, the elasticity governs finite changes in logs: halving tightness multiplies unemployment by $\sqrt{2}$. At the 2022:Q2 peak, tightness was 2 and unemployment was 3.5\%, so the FERU was $u^*=\sqrt{uv}=5.0\%$. Cooling tightness from 2 to 1 along that Beveridge curve therefore takes unemployment to the FERU, from 3.5\% to 5.0\%. The rise is 1.5pp---not zero, but far from a recession. A soft landing from an inefficiently tight labor market is therefore possible, because the labor market starts on the steep portion of the aggregate supply curve.

The actual path was even softer. By July 2025, tightness had returned to 0.99 but unemployment stood at only 4.3\% \citep{ABCD}. That extra 0.7pp never materialized thanks to the inward Beveridge shift visible in figure~\ref{f:feru}. After the pandemic outward jump, the FERU was still falling. The unwinding of the pandemic matching shock reduced the unemployment cost of inflation cooling.

\subsection{Choice of target variables} 

Since the two mandates coincide, targeting the inflation rate or targeting the FERU is equivalent in the model. In practice, however, the Fed should target the variable that is most volatile---to observe most clearly departures from the dual mandate.

The slope of the Phillips curve determines which variable is most volatile. If the Phillips curve is steep, demand shocks mostly move inflation, so inflation is the more informative target. If the Phillips curve is flat, demand shocks mostly move unemployment, so unemployment is the more informative target. Because the Phillips curve is convex---flat when unemployment is inefficiently high, steep when it is inefficiently low---the implication is state-contingent: the Fed should target the FERU when the economy is inefficiently slack, and target 2\% inflation when the economy is inefficiently tight. Since inflation and tightness vary commensurately (the tightness-inflation curve is broadly linear), an equivalent prescription when tightness is easier to measure than inflation is to target the FERU in slumps and the efficient tightness $\t^* = 1$ in booms.

\subsection{The NAIRU is not well defined}\label{s:nairu}

The NAIRU is the unemployment rate at which inflation is stable; it is estimated from accelerationist Phillips curves. In the model, the NAIRU is simply not well defined: inflation is stable at any unemployment rate. If unemployment is at the FERU, inflation is stable at the target; if unemployment is above the FERU, inflation is stable below the target; if unemployment is below the FERU, inflation is stable above the target. The Phillips curve is not accelerationist. This may explain why NAIRU estimates are notoriously imprecise \citep{SSW97a,L01}.

\subsection{Ineffective monetary communication}\label{s:communication}

The divine coincidence rests on the assumption that the central bank anchors normal inflation at its target: $\pi^n = \pi^*$. If monetary communication is ineffective, so that $\pi^n \neq \pi^*$, the derivations are unchanged except that $\pi^*$ is replaced by $\pi^n$ in the Phillips curve \eqref{e:pc}. It is as if a shifter $\pi^n - \pi^*$ were added to the curve:
\begin{equation}
\pi = (\pi^n - \pi^*) + \pi^* + \frac{y(\t)}{\vp \d (\d-r)} \bs{\frac{\h(\t)}{1-\h(\t)} \cdot \frac{\tau(\t)}{u(\t)} - 1}.
\label{e:nodivine}\end{equation}
At full employment, inflation is now $\pi^n$ instead of $\pi^*$: the divine coincidence breaks down, and the two mandates cannot be achieved simultaneously. If people believe that inflation will normally run above target ($\pi^n>\pi^*$), then inflation is too high when the economy is at full employment; equivalently, when inflation is on target, the labor market is inefficiently slack. Hence, in this model, the coincidence is not so much ``divine'' as the product of careful and effective messaging by the central bank. Such messaging appears to have succeeded in the United States since the mid-1990s. Long-run inflation expectations have remained glued to 2\% \citep{HHN22}, explaining why the coincidence is visible in modern US data.

\section{Conclusion}\label{s:conclusion}

This paper develops a Beveridgean model of the Phillips curve. The economy is a matching economy in which households purchase services from other households, and all trades are mediated by a matching function---generating unemployment, vacancies, and a Beveridge curve. Directed search generates price competition: households post prices for the services they buy, and jobseekers arbitrage across submarkets. Price-adjustment costs generate price rigidity. Combining the two produces a nonvertical Phillips curve anchored at the efficient allocation.

The model explains the three facts that the post-pandemic economy revealed about the US Phillips curve. Inflation is on target when unemployment is at the FERU, because directed search is efficient and the Fed anchors normal inflation at its target. Slack drives inflation, and the tightness formulation of the Phillips curve is more stable than the unemployment formulation, because the efficient tightness is pinned down by two stable statistics while the FERU moves with the Beveridge curve. And the Phillips curve is convex in the unemployment-inflation plane: steep when the labor market is inefficiently tight, flat when it is inefficiently slack. The convexity comes from the hyperbolic shape of the Beveridge curve. It explains why the drop in inflation was so muted when unemployment spiked in 2020, and why inflation flared up so quickly when the labor market overheated in 2021--2023.

Beyond the Phillips curve, Moen-Rotemberg pricing could be useful in other matching models. Matching models of any market or economy require some price rigidity to generate realistic fluctuations in tightness \citep{S05,MS15}. Existing mechanisms rely on random search: fixed price norms \citep{H05}; credible bargaining \citep{HM08}; staggered Nash bargaining \citep{GT09}; and rigid price norms \citep{BG10,M12,M24,MS15,MS19}. Moen-Rotemberg pricing offers an alternative built on directed search, where the degree of rigidity is disciplined by measurable price-adjustment costs.

\bibliography{\bib}
\newpage
\appendix

\section{Convexity of the Phillips curve}\label{a:convexity}

This appendix proves proposition~\ref{r:convexity}: the Phillips curve $\pi(u)$ is strictly convex in $u$ under the assumption that the matching function is Cobb-Douglas and symmetric,
\begin{equation*}
m(u,v) = \m\sqrt{uv}. 
\end{equation*}
It is difficult to obtain more general analytical results; for other matching functions, numerical analysis is required, as in section~\ref{s:quantitative}.

\subsection{Symmetric Cobb-Douglas matching function}

With a symmetric Cobb-Douglas matching function, the trading rates are 
\begin{equation}
f(\t)=\m \sqrt{\t} \quad\text{and}\quad q(\t)=\frac{\m}{\sqrt{\t}},
\label{e:fq}\end{equation}
and the matching elasticity is constant at $\h(\t)=1/2$, so 
\begin{equation}
\frac{\h(\t)}{1-\h(\t)} = 1.
\label{e:eta}\end{equation}

The unemployment rate is given by \eqref{e:u}, so
\begin{equation*}
u(\t) = \frac{\l}{\l + \m \sqrt{\t}}.
\end{equation*}
Inverting this strictly decreasing relationship, we write tightness as a strictly decreasing function of the unemployment rate:
\begin{equation}
\t(u) = \bs{\frac{\l (1-u)}{\m u}}^2.
\label{e:thetau}\end{equation}

The highest possible value for tightness is $\t\to \bar{\t}$, when $\tau(\t)\to \infty$. The upper bound $\bar{\t}$ is defined such that $q(\bar{\t}) = \k\l$, so here
\begin{equation*}
\bar{\t} = \bp{\frac{\m}{\k\l}}^2.
\end{equation*}
The lowest possible unemployment rate is achieved when $\t\to \bar{\t}$. We denote this minimum unemployment rate by $\bar{u} = u(\bar{\t})$; it is given by
\begin{equation}
\bar{u} = \frac{\k \l^2}{\m^2 + \k \l^2}.
\label{e:ubar}\end{equation}

\subsection{Preliminary result}

\begin{lemma}\label{r:ratio}
The ratio between the matching wedge and unemployment rate admits the closed form
\begin{equation}
\frac{\tau(u)}{u} = \frac{1-\bar{u}}{u-\bar{u}} - \frac{1}{u}.
\label{e:ratio}
\end{equation}
The ratio is strictly decreasing and strictly convex in $u$ on $(\bar{u},1)$.
\end{lemma}

The proof has three steps: derive the closed form \eqref{e:ratio}, then sign the first and second derivatives.

First, combining \eqref{e:fq} and \eqref{e:thetau} allows us to express the ad-filling rate as a function of the unemployment rate:
\begin{equation*}
q(u) = \frac{\m^2}{\l}\cdot\frac{u}{1-u}.
\end{equation*}
Using expression \eqref{e:tau}, we can express the matching wedge as a function of the unemployment rate too:
\begin{equation*}
\tau(u) = \frac{\k \l}{\dfrac{\m^2}{\l}\dfrac{u}{1-u} - \k\l} = \frac{\k \l^2(1-u)}{(\m^2+\k\l^2) u - \k\l^2} = \frac{\bar{u}(1-u)}{ u - \bar{u}},
\end{equation*}
where the last equality uses the definition \eqref{e:ubar} of $\bar{u}$. Then, the ratio between matching wedge and unemployment rate is a simple function of the unemployment rate:
\begin{equation*}
\frac{\tau(u)}{u} = \frac{\bar{u}(1-u)}{u (u - \bar{u})} = \frac{\bar{u} - u + u - \bar{u}u}{u (u - \bar{u})} = \frac{1-\bar{u}}{u-\bar{u}} - \frac{1}{u},
\end{equation*}
which establishes \eqref{e:ratio}.

Second, differentiating \eqref{e:ratio} once with respect to the unemployment rate, we get
\begin{equation}
\od{\tau/u}{u} = \frac{1}{u^2} -\frac{1-\bar{u}}{(u-\bar{u})^2}.
\label{e:first}\end{equation}
The sign of the derivative is the sign of
\begin{equation*}
\bp{u - \bar{u}}^2 - (1-\bar{u}) u^2 = \bar{u} \bs{\bar{u} - 2 u + u^2} = \bar{u} \bs{(\bar{u}-u) + u (u-1)}.
\end{equation*}
Since $0<\bar{u}<u<1$, we have $u-1<0$ and $\bar{u}-u<0$, so the first derivative is strictly negative: $\tau/u$ is strictly decreasing in $u$.

Third, differentiating \eqref{e:first} with respect to the unemployment rate, we get the second derivative of the ratio:
\begin{equation*}
\od[2]{\tau/u}{u} = \frac{2(1-\bar{u})}{(u-\bar{u})^3} - \frac{2}{u^3}.
\end{equation*}
The sign of the second derivative is the sign of
\begin{align*}
(1-\bar{u}) u^3 - (u-\bar{u})^3 & = \bar{u} \bs{- u^3 + \bar{u}^2 + 3 u^2 - 3 u \bar{u}}\\
& = \bar{u} \bs{u^2 (1-u) + (\bar{u} - u)^2 + u (u - \bar{u})}.
\end{align*}
Since $0<\bar{u}<u<1$, all the terms in the square bracket are positive: $u^2 (1-u)>0$, $(\bar{u} - u)^2>0$, and $u (u - \bar{u})>0$. Hence the second derivative is strictly positive: $\tau(u)/u$ is strictly convex in $u$.

\subsection{Proof of proposition~\ref{r:convexity}}

With lemma~\ref{r:ratio} in hand, we can now prove the proposition. 

The Phillips curve $\pi(\t)$ is given by \eqref{e:pc}, where here \eqref{e:eta} holds. We compose it with the function $\t(u)$ given by \eqref{e:thetau} to obtain a Phillips curve linking inflation to unemployment:
\begin{equation*}
\pi(u) = \pi^* + \frac{1}{\vp \d (\d-r)} \cdot y(u) \cdot \bs{\frac{\tau(u)}{u} - 1}.
\end{equation*}
The ratio $\tau(u)/u$ is given by \eqref{e:ratio} and output is given by $y(u) = (1-u)kh$. Given the linear expression for output, we can rewrite the Phillips curve as
\begin{equation}
\pi(u) = \pi^* + \frac{kh}{\vp \d (\d-r)} \cdot (1-u) \cdot \bs{\frac{\tau(u)}{u} - 1}.
\label{e:pcu}\end{equation}

We differentiate \eqref{e:pcu} twice. The first derivative is
\begin{equation*}
\pi'(u) = \frac{kh}{\vp \d (\d-r)} \cdot \bs{(1-u) \cdot \od{\tau/u}{u} + 1 - \frac{\tau(u)}{u}}.
\end{equation*}
At efficiency, by \eqref{e:efficiency} and \eqref{e:eta}, $\tau(u^*)/u^* = 1$, so the sign of $\pi'(u^*)$ is the sign of $(1-u^*) \odx{(\tau/u)}{u}$. By lemma~\ref{r:ratio}, $\odx{(\tau/u)}{u}<0$, so $\pi'(u^*)<0$: in the vicinity of the FERU, the Phillips curve is strictly decreasing in the unemployment rate.

Differentiating again, we obtain
\begin{equation*}
\pi''(u) = \frac{kh}{\vp \d (\d-r)} \cdot \bs{(1-u) \cdot \od[2]{\tau/u}{u} - 2 \cdot \od{\tau/u}{u}}.
\end{equation*}
By lemma~\ref{r:ratio}, $\odx{(\tau/u)}{u}<0$ and $\odx[2]{(\tau/u)}{u}>0$, so both terms in the square brackets are strictly positive, and $\pi''(u)>0$ on $(\bar{u},1)$: the Phillips curve is strictly convex in the unemployment rate.

\section{Dynamical system at the zero lower bound}\label{a:zlb}

This appendix derives and studies the dynamical system describing the behavior of tightness and inflation at the zero lower bound.

\subsection{Costate variables, inflation, and tightness}

The first step is to express the costate variables $\p^b$ and $\p^p$ as functions of inflation $\pi$ and tightness $\t$. Using these relationships, we transform the differential equations for the costate variables into differential equations for inflation and tightness.

For the costate variable $\p^p$, the relationship is directly given by \eqref{e:foc2s}:
\begin{equation}
\p^p = \vp \bs{\pi-\pi^*},\qquad \dot{\p}^p = \vp \dot{\pi}.
\label{e:psip}\end{equation}

For the costate variable $\p^b$, the relationship arises from \eqref{e:foc1s}. In the model, aggregate consumption is directly determined by tightness, from \eqref{e:c}:
\begin{equation}
c(\t) = \frac{1-u(\t)}{1+\tau(\t)} \cdot kh.
\label{e:czlb}\end{equation}
Notice also that because $y(\t) = \bs{1 + \tau(\t)} c(\t)$, condition \eqref{e:foc1s} relates $\p^b$ to tightness:
\begin{equation}
\p^b(\t) = \frac{a}{\sqrt{[1 + \tau(\t)] y(\t)}}.
\label{e:psib2}\end{equation}
This equation implies:
\begin{equation}
\p^b(\t) y(\t) = a\sqrt{c(\t)}.
\label{e:psiby}\end{equation}

By the chain rule, $\dot{\p}^b = (\p^b/\t) \cdot \e^{\p^b}_{\t} \cdot \dot{\t}$, so we need the elasticity of $\p^b$ with respect to tightness. From \eqref{e:tauelasticity}, $\e^{1+\tau}_{\t} = \h(\t) \tau(\t)$; from \eqref{e:uelasticity}, $\e^{1-u}_{\t} = \bs{1-\h(\t)} u(\t)$. Hence, from \eqref{e:czlb}, the elasticity of consumption is 
\begin{equation*}
\e^c_{\t} = \bs{1-\h(\t)} u(\t) - \h(\t) \tau(\t),
\end{equation*}
and from \eqref{e:psib2}, the elasticity of the costate variable is
\begin{equation*}
\e^{\p^b}_{\t} = -\frac{1}{2}\cdot\bs{\bs{1-\h(\t)} u(\t) + \h(\t) \tau(\t)}.
\end{equation*}
The time derivative of the costate variable is therefore related to that of tightness by
\begin{equation}
\frac{\dot{\p}^b}{\p^b} = -\frac{1}{2}\cdot \bs{\bs{1-\h(\t)} u(\t) +\h(\t) \tau(\t)} \cdot\frac{\dot{\t}}{\t}.
\label{e:psibdot}\end{equation}

\subsection{Differential equations for inflation and tightness}

We first transform the differential equation for the costate variable $\p^p$, given by \eqref{e:foc4s}, into a differential equation for inflation. We replace $\p^p$ and $\dot{\p}^p$ by the expressions \eqref{e:psip}, and $\p^b y$ by the expression \eqref{e:psiby}. We obtain equation \eqref{e:pidot}:
\begin{equation*}
\dot{\pi} = \d \bs{\pi-\pi^*} + \frac{a}{\vp} \cdot \sqrt{c(\t)} \bs{1 -\frac{\h(\t)}{1-\h(\t)} \cdot \frac{\tau(\t)}{u(\t)}}.
\end{equation*}
The nullcline $\dot{\pi} = 0$ is given by \eqref{e:pi0}. This nullcline is close to, but not the same as, the Phillips curve in normal times \eqref{e:pc}: in normal times the costate variable $\p^b$ is fixed by the real interest rate, whereas at the zero lower bound it is jointly determined by tightness and inflation.

Next, we transform the differential equation for the costate variable $\p^b$, given by \eqref{e:foc3s}, into a differential equation for tightness. We divide the equation by $\p^b$ and set the real interest rate to $r = -\pi$:
\begin{equation*}
\frac{\dot{\p}^b}{\p^b} = \bp{\d + \pi} - \frac{1}{\p^b}.
\end{equation*}
Replacing $\p^b$ and $\dot{\p}^b$ by the expressions \eqref{e:psib2} and \eqref{e:psibdot} yields equation \eqref{e:thetadot}:
\begin{equation*}
\frac{1}{2}\cdot \bs{\bs{1-\h(\t)} u(\t) + \h(\t) \tau(\t)} \cdot\frac{\dot{\t}}{\t} = \frac{1}{a} \cdot \sqrt{\bs{1 + \tau(\t)} y(\t)} - \bp{\d + \pi}.
\end{equation*}
The nullcline $\dot{\t} = 0$ is given by \eqref{e:theta0}.

\subsection{Equilibrium of the dynamical system}

We now determine the equilibrium of the dynamical system composed of \eqref{e:thetadot} and \eqref{e:pidot}. At the equilibrium, $\dot{\pi} = \dot{\t} = 0$, so the equilibrium lies at the intersection of the two nullclines. Let us introduce auxiliary functions to simplify the analysis:
\begin{align*}
P(\t) &= \pi^* + \frac{a}{\d\vp} \cdot \sqrt{c(\t)} \bs{\frac{\h(\t)}{1-\h(\t)} \cdot \frac{\tau(\t)}{u(\t)} - 1}\\
Q(\t) &= \frac{1}{a} \cdot \sqrt{\bs{1 + \tau(\t)} y(\t)} - \d\\
R(\t)&= \frac{2\t}{\bs{1-\h(\t)} u(\t) +\h(\t) \tau(\t)}.
\end{align*}
The nullcline \eqref{e:pi0} is $\pi = P(\t)$ and the nullcline \eqref{e:theta0} is $\pi = Q(\t)$. Let $S(\t) = Q(\t) - P(\t)$: the tightness that solves the model satisfies $S(\t) = 0$.

At $\t=0$, we have $u(0)=1$ so $c(0) = 0$ and $S(0) = - (\d+\pi^*)<0$. At $\t = \t^*$, we have $\tau/u = \bs{1-\h}/\h$ by \eqref{e:efficiency}, so
\begin{equation*}
S(\t^*) = \frac{1}{a} \cdot \sqrt{\bs{1 + \tau(\t^*)} y(\t^*)} - (\d+\pi^*).
\end{equation*}
The economy is at the zero lower bound only when aggregate demand is depressed enough, which occurs when the taste for consumption satisfies \eqref{e:zlbcondition}---and \eqref{e:zlbcondition} is equivalent to $S(\t^*)>0$. We have established that $S(0)<0$ and $S(\t^*)>0$, and $S$ is continuous on $[0,\t^*]$; by the intermediate value theorem, the equation $S(\t)=0$ admits a solution on $(0,\t^*)$. The dynamical system therefore admits an equilibrium on $(0,\t^*)$, which we denote $(\t^z,\pi^z)$.

Is the equilibrium unique? It would be if $S$ were strictly increasing on $(0,\t^*)$. The function $Q$ is strictly increasing on $(0,\t^*)$, as $\tau(\t)$ and $y(\t)$ are increasing on the interval. But the function $P$ is not monotone (indeed $P(0) = P(\t^*) = \pi^*$), so we cannot conclude in general. However, when prices are rigid enough---when $\vp$ is large enough---the fluctuations in $P$ become arbitrarily small: at the limit $\vp \to \infty$, $P(\t) \to \pi^*$ for any $\t$. So for $\vp$ large enough, $S$ never returns to 0 once it has crossed 0 at $\t^z$, and the equilibrium is unique. The equilibrium is unique under the calibration of figure~\ref{f:zlbphase}. In that case, since $S(0)<0$ and $S(\t^*)>0$, $S$ crosses 0 from below, so $S'(\t^z)>0$. In the rest of the appendix we assume that price rigidity $\vp$ is sufficient to guarantee uniqueness.

\subsection{Classification of the dynamical system}

Finally, we determine whether the dynamical system is a source, sink, or saddle around the zero-lower-bound equilibrium, by computing the Jacobian of the system, which describes the local dynamics \citep[pp.~61--64]{HSD13}.

With the auxiliary functions, the differential equation \eqref{e:thetadot} is $\dot{\t} = R(\t) \bs{Q(\t) -\pi}$ and the differential equation \eqref{e:pidot} is $\dot{\pi} = \d \bs{\pi - P(\t)}$. The Jacobian of the system $[\pi(t), \t(t)]$ is
\begin{equation*}
\J = \bs{\begin{array}{cc}
\d & -\d P'(\t^z) \\ 
-R(\t^z) & R(\t^z) Q'(\t^z)\\ 
\end{array}},
\end{equation*}
where we simplified the Jacobian using $Q(\t^z) -\pi^z =0$, which holds by definition of the equilibrium. The trace and determinant of the Jacobian are
\begin{align*}
\tr(\J) &= \d + R(\t^z) Q'(\t^z)\\ 
\det(\J) &= \d R(\t^z) \bs{Q'(\t^z) - P'(\t^z)} = \d R(\t^z) S'(\t^z).
\end{align*}
We know that $\d>0$, $R(\t^z)>0$, and $Q'(\t^z)>0$, so the trace is strictly positive. We also know that $S'(\t^z)>0$, so the determinant is strictly positive. Thus, the dynamical system describing inflation and tightness is a source around the zero-lower-bound equilibrium.

\end{document}